\documentclass[aps,physrev,preprint,groupedaddress,showkeys]{revtex4-2}

\usepackage{graphicx}
\usepackage{newtxtext}
\usepackage{newtxmath}
\usepackage{natbib}
\usepackage{hyperref}
\usepackage{subfigure}
\usepackage{overpic}
\usepackage{xcolor}
\hypersetup{
    colorlinks = true,
    urlcolor   = blue,
    citecolor  = black,
}

\newcommand{\RomanNumeralCaps}[1]
\linenumbers

\newcommand{\rbr}[1]{\left( #1 \right)}
\newcommand{\sbr}[1]{\left[ #1 \right]}
\newcommand{\abr}[1]{\left\langle #1 \right\rangle}

\begin{document}

\title{Area rule of velocity circulation in two-dimensional instability-driven turbulence beyond the inertial range}

\author{Bo-Jie Xie}
\affiliation{College of Mechanics and Engineering Science and State Key Laboratory for Turbulence and Complex Systems, Peking University, Beijing 100871, People’s Republic of China}
\author{Tian-Shu Zhou}
\affiliation{College of Mechanics and Engineering Science and State Key Laboratory for Turbulence and Complex Systems, Peking University, Beijing 100871, People’s Republic of China}
\author{Jin-Han Xie}
\email{jinhanxie@pku.edu.cn}
\affiliation{College of Mechanics and Engineering Science and State Key Laboratory for Turbulence and Complex Systems, Peking University, Beijing 100871, People’s Republic of China}

\begin{abstract}
The velocity statistics reveal non-universality in both three-dimensional (3-D) and two-dimensional (2-D) turbulence, despite both prototype systems containing an energy inertial range with constant energy flux.
Recently, statistics of scale-dependent velocity circulation exhibit universal bifractal behavior in 2-D and 3-D hydrodynamic turbulence and quantum turbulence, where the circulation scale is defined as the square root of the minimum area enclosed by the loop. 
This loop-shape independent definition of scale bases on the area rule of circulation first proposed by Migdal: the probability density function (PDF) of circulation is only a function of the minimal surface area enclosed by the loop but not the shape of the loop.
This paper demonstrates that the derivation of the circulation area rule can be generalized to all scales in 2-D instability-driven turbulence, not limited to the inertial range. 
However, the area rule is not the only solution to the loop equation, so it may not be observed.
Another necessary condition for the validity of the area rule is that the second-order momentum of circulation is loop-shape independent.
By deriving the relationship between the second-order moment of circulation on a rectangular loop and the energy spectrum, we prove that the area rule cannot be satisfied in the classic inertial-range turbulence with $-5/3$ or $-3$ spectral scalings.
As in the 3-D case \citep{Iyer2021}, the second-order moment of circulation is size-dependent. Compared with the circulation PDFs, the PDFs normalized by the second-order moment of circulation exhibit significantly weaker dependence on loop shape.
\end{abstract}


\maketitle

\section{Introduction}
\label{sec:headings}

Since Kolmogorov's theory for homogeneous isotropic three-dimensional turbulence \cite{kolmogorov1941local}, structure functions have long been used to study the statistical properties of turbulence in and out of the inertial range where the energy flux is scale-independent.
However, the intermittent nature of turbulence prevents a simple understanding of high-order statistics \cite{Kolmogorov1962,Oboukhov1962,Anselmet1984,Gagne1987}, and theories were proposed to capture intermittency \citep{Benzi1993,She1994,Frisch1995,Sreenivasan1997}. 
As to the two-dimensional (2-D) turbulence \citep{Kraichnan1967}, due to the conservation of energy and enstrophy, a dual cascade scenario with the downscale transfer of enstrophy and upscale energy transfers exists, which distinguishes it from three-dimensional (3-D) turbulence \citep{Paret1998}.
Therefore, seeking universality in both two- and three-dimensional turbulence is highly nontrivial.

Velocity circulation has recently been proposed to explore universality.
\cite{Migdal1995} proposed the theory of circulation statistics in the early 1990s with an area rule stating that the PDF of velocity circulation depends on the minimum area enclosed by the loop. 
\cite{Makoto1993} and \cite{Cao1996} pointed out that as the area enclosed by the loop increases, the PDF of the loop tends towards a Gaussian distribution. 
Recently, \cite{Iyer2019} validated Migdal's area rule for simple loops using high-precision direct numerical data and affirmed that for an $8$-loop, the PDF of velocity circulation depends on the scalar area enclosed by the loop. 
They also pointed out that velocity circulation is bifractal, which is simpler than the multifractal feature of velocity structure function. Afterwards, this bifractal property of velocity circulation statistics was also found in quantum turbulence \citep{Muller2021,Polanco2021} and the energy inertial range of 2-D turbulence \citep{Zhu2023}, making velocity circulation a promising quantity to study turbulence universality. In 2-D flow, high-order statistics of circulation in classical and quantum turbulence also exhibit similar properties \citep{Muller_Krstulovic_2024}. The self-similarity breaking of circulation is also observed in 2-D turbulence with linear damping \citep{Muller_Krstulovic_2025}.

With many exciting results in circulation statistics, most studies emphasize the high-order moments of the circulation in inertial ranges, while the applicability of the area rule has not received sufficient attention. 
Considering the opposite energy transfer directions in 3-D turbulence and 2-D turbulence, we expect distinct properties of these two systems. So the applicability of the area rule in the 2-D case is an important question. In addition, whether the area rule holds when the loop is not in the inertial range remains another question. 
Although \cite{Zhu2023} studied the PDFs of velocity circulation on the loops crossing the energy and enstrophy inertial ranges and enclosing the same area, and found that these PDFs clearly did not coincide, which means the area rule failed, we cannot determine whether the failure of the area rule was caused by non-conservative forcing or by the energy transfer properties of 2-D turbulence. So, for self-driven turbulence without non-conservative forcing, the applicability of the area rule remains to be debated.
Also, we show that the area rule fails in classic turbulence systems with inertial-range spectral exponents of $-5/3$ and $-3$. Therefore, this paper focuses on scales beyond the inertial range, where energy injection and dissipation are important.
To capture the effects of energy injection and dissipation with a representative but straightforward form, we study 2-D instability-driven turbulence.

The structure of this paper is as follows. 
\S \ref{Sec_2ndmoment} presents the exact relation between energy spectrum and second-order moment of velocity circulation for rectangular loops, and therefore, the inertial-range turbulence does not support the area rule, which motivates our derivation of the area rule for instability-driven 2-D turbulence in \S \ref{Sec_derivation}.
\S \ref{Sec_numerics} numerically tests the area rule in simple and complex loops using rectangular loops and 8-loops as examples.
We summarize and discuss our key results in \S \ref{Sec_diss}.

\section{Relationship between the second-order moment of velocity circulation and energy spectrum}
\label{Sec_2ndmoment}

In two dimensions, velocity circulation on an orientable loop $C$ is defined as
\begin{equation}\label{2D vorticity and velocity circulation}
    \Gamma _C = \oint _C \boldsymbol{v} \cdot d\boldsymbol{r} = \iint _S \omega d\sigma ,
\end{equation}
where $\boldsymbol{v} = u \boldsymbol{e}_x + v \boldsymbol{e}_y$ is velocity, $d\boldsymbol{r}$ is line element on the loop $C$, $\omega= \partial_x v - \partial_y u$ is vorticity, $S$ is the area encompassed by the loop $C$, $d \sigma = dx \wedge dy$ is area element on $S$. The second equality uses the Stokes theorem.

Let $G(x)$ be a function defined in a periodic box $D$ of size $L_x \times L_y$, where $L_x$ and $L_y$ are the lengths of the edges of $D$ parallel to the $x$ and $y$ directions, respectively. $\omega (\boldsymbol{x})$ is a vorticity field in this periodic box. 
Considering a filtered vorticity
\begin{equation}
    \tilde{\Gamma}(\boldsymbol{x}) \equiv \int_{D}{\omega (\boldsymbol{x}') G (\boldsymbol{x} - \boldsymbol{x}') d\boldsymbol{x}'} ,
    \label{Gamma field}
\end{equation}
because the function $G(\boldsymbol{x})$ and the vorticity field $\omega (\boldsymbol{x})$ are periodic, we can use Fourier series to express them: 
\begin{subequations}\label{Fourier series}
    \begin{align}
        G(\boldsymbol{x}) &= \sum_{k = -\infty}^{\infty}{\sum_{l = -\infty}^{\infty}{\hat{G}_{k,l} e^{ik\frac{2 \pi}{L_x}x} e^{il\frac{2 \pi}{L_y}y}}} , \\
        \omega (\boldsymbol{x}) &= \sum_{k = -\infty}^{\infty}{\sum_{l = -\infty}^{\infty}{\hat{\omega}_{k,l} e^{ik\frac{2 \pi}{L_x}x} e^{il\frac{2 \pi}{L_y}y}}} ,
    \end{align}
\end{subequations}
where
\begin{subequations}\label{coefficients of Fourier series}
    \begin{align}
        \hat{G}_{k,l} &= \frac{1}{L_x L_y} \int_D {G (\boldsymbol{x}) e^{-ik\frac{2 \pi}{L_x}x} e^{-il\frac{2 \pi}{L_y}y} dxdy} , \\
        \hat{\omega}_{k,l} &= \frac{1}{L_x L_y} \int_D {\omega (\boldsymbol{x}) e^{-ik\frac{2 \pi}{L_x}x} e^{-il\frac{2 \pi}{L_y}y} dxdy} . 
    \end{align}
\end{subequations}
Substitute (\ref{Fourier series}) into (\ref{Gamma field}) while using
\begin{equation}
    \int_{D} {e^{i(k-k')\frac{2 \pi}{L_x}x} e^{i(l-l')\frac{2 \pi}{L_y}y} dxdy} = L_x L_y \delta_{k,k'} \delta_{l,l'} 
\end{equation}
where 
\begin{equation}
    \delta_{k,k'} = 
    \begin{cases}
        0& (k \neq k')\\
        1& (k = k')
    \end{cases}
\end{equation}
is the Kronecker delta, we can get
\begin{equation}\label{Fourier series of Gamma}
    \tilde{\Gamma}(\boldsymbol{x}) = L_x L_y \sum_{k = -\infty}^{\infty}{\sum_{l = -\infty}^{\infty} {\hat{\omega}_{k,l} \hat{G}_{k,l} e^{ik\frac{2 \pi}{L_x}x} e^{il\frac{2 \pi}{L_y}y}}} .
\end{equation}

For homogeneous flow field, when $G(\boldsymbol{x})$ is given, $\abr{\tilde{\Gamma}(\boldsymbol{x})^2}$ should not change with the variation of $\boldsymbol{x}$, so 
\begin{equation}\label{ensemble average equals to spacial average}
    \abr{\tilde{\Gamma}(\boldsymbol{x})^2} = \frac{1}{L_x L_y} \int_{D} {\abr{\tilde{\Gamma}(\boldsymbol{x})^2} d\boldsymbol{x}} .
\end{equation}
Since $\tilde{\Gamma}$ must be real, $\tilde{\Gamma}^2 = \tilde{\Gamma} \tilde{\Gamma}^*$, where $\tilde{\Gamma}^*$ is conjugate of $\tilde{\Gamma}$. Then substitute (\ref{Fourier series of Gamma}) into (\ref{ensemble average equals to spacial average}), we can get
\begin{equation}\label{relationship between second-order moment of circulation and enstrophy spectrum}
    \abr{\tilde{\Gamma}(\boldsymbol{x})^2} = L_x^2 L_y^2 \sum_{k = -\infty}^{\infty} \sum_{l = -\infty}^{\infty} {\abr{|\hat{\omega}_{k,l}|^2} |\hat{G}_{k,l}|^2} ,
\end{equation}
where $\abr{|\hat{\omega}_{k,l}|^2}$ is the spectrum of enstrophy.

Let $u(\boldsymbol{x})$ and $v(\boldsymbol{x})$ be the $x$-direction and $y$-direction components of the velocity, respectively. For incompressible fluids, 
\begin{equation}\label{u hat and v hat}
    \frac{k}{L_x} \hat{u}_{k,l} = - \frac{l}{L_y} \hat{v}_{k,l}
\end{equation}
can be obtained based on incompressible conditions, where $\hat{u}_{k,l}$ and $\hat{v}_{k,l}$ are the coefficients of the Fourier series of $u(\boldsymbol{x})$ and $v(\boldsymbol{x})$, respectively. Because $\omega = \partial_x v - \partial_y u$, we can use (\ref{u hat and v hat}) to rewrite (\ref{relationship between second-order moment of circulation and enstrophy spectrum}) as
\begin{equation}\label{second-order moment and energy spectrum}
    \abr{\tilde{\Gamma}(\boldsymbol{x})^2} = L_x^2 L_y^2 \sum_{k = -\infty}^{\infty} \sum_{l = -\infty}^{\infty} {\sbr{k^2 \rbr{\frac{2\pi}{L_x}}^2 + l^2 \rbr{\frac{2\pi}{L_y}}^2} \hat{E}_{k,l} |\hat{G}_{k,l}|^2} ,
\end{equation}
where $\hat{E}_{k,l} \equiv \abr{|\hat{u}_{k,l}|^2 + |\hat{v}_{k,l}|^2}$ is the spectrum of energy.
Especially, when $G(\boldsymbol{x}) = H \rbr{\frac{a}{2} - |x + \frac{a}{2}|}$ $H \rbr{\frac{b}{2} - |y + \frac{b}{2}|}$, where $H(x)$ is Heaviside function, $(x,y) \in D$ and $[-a,0) \times [-b,0) \subseteq D$, $\tilde{\Gamma} (\boldsymbol{x})$ is the circulation $\Gamma (\boldsymbol{x})$ on a rectangular loop with length $a$ and width $b$ and four vertices of $(x,y)$, $(x+a,y)$, $(x+a,y+b)$ and $(x,y+b)$. So for any incompressible flow field in a periodic box, the second-order moment of the circulation on any rectangular loop $C$ with length $a$ and width $b$ ($a < L_x$, $b < L_y$) can be calculated using the following formula:
\begin{equation}\label{circulation second-order moment and energy spectrum}
    \abr{\Gamma_C^2} = 4 \sum_{k = -\infty}^{\infty} \sum_{l = -\infty}^{\infty} {\sbr{ \rbr{\frac{L_y}{\pi l}}^2 + \rbr{\frac{L_x}{\pi k}}^2} \sin^2{\rbr{\frac{\pi k}{L_x} a}} \sin^2{\rbr{\frac{\pi l}{L_y} b}} \hat{E}_{k,l}} .
\end{equation}

\subsection{Implication on area rule}

Suppose the area rule holds in the energy or enstrophy inertial range of 2-D turbulence. Then, a necessary condition is that, for loops with the same area but varying shape, the second-order moment of the velocity circulation is invariant; e.g., for rectangular loops with different aspect ratios, the second-order moment of the velocity circulation does not vary with the aspect ratio. 

\begin{figure}
    \centering
    \begin{overpic}
    [width=0.45\textwidth]{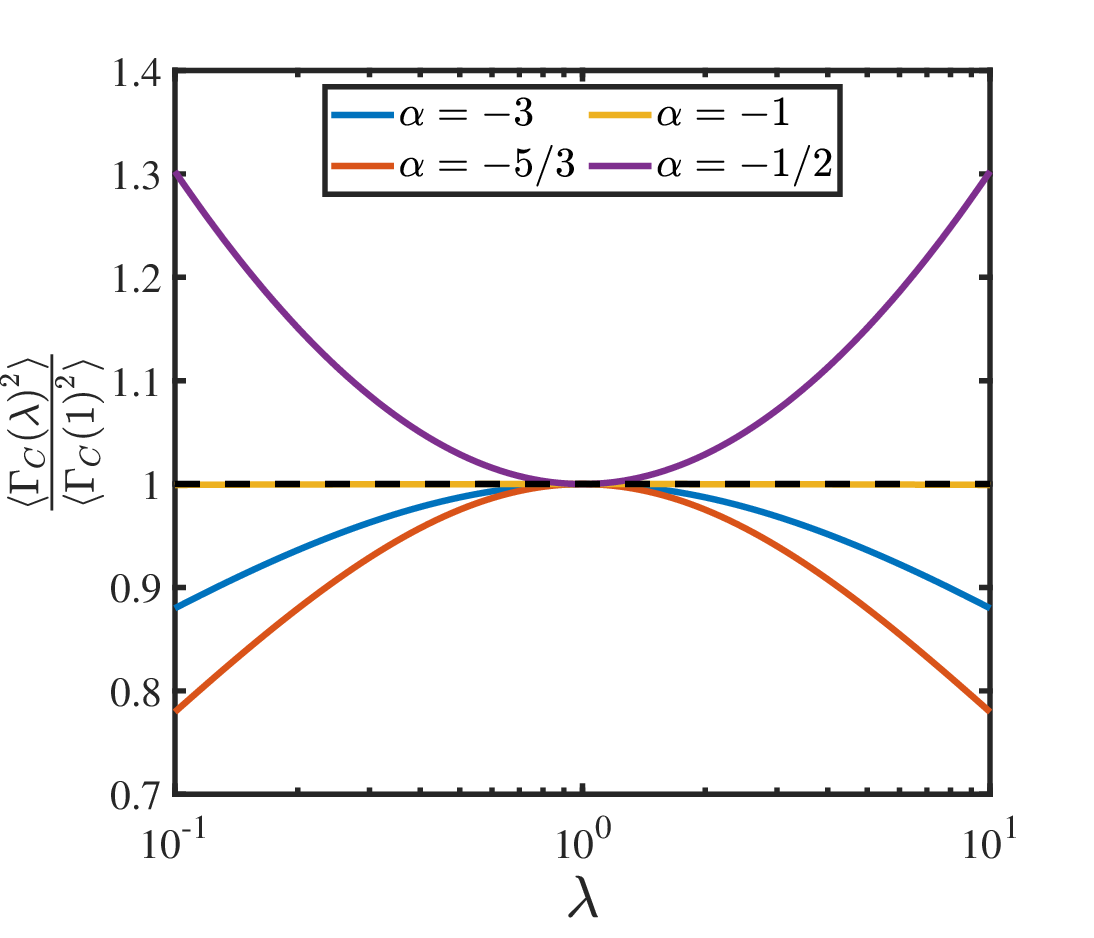}
        \put(0,80){(a)}
    \end{overpic}
    \begin{overpic}
    [width=0.45\textwidth]{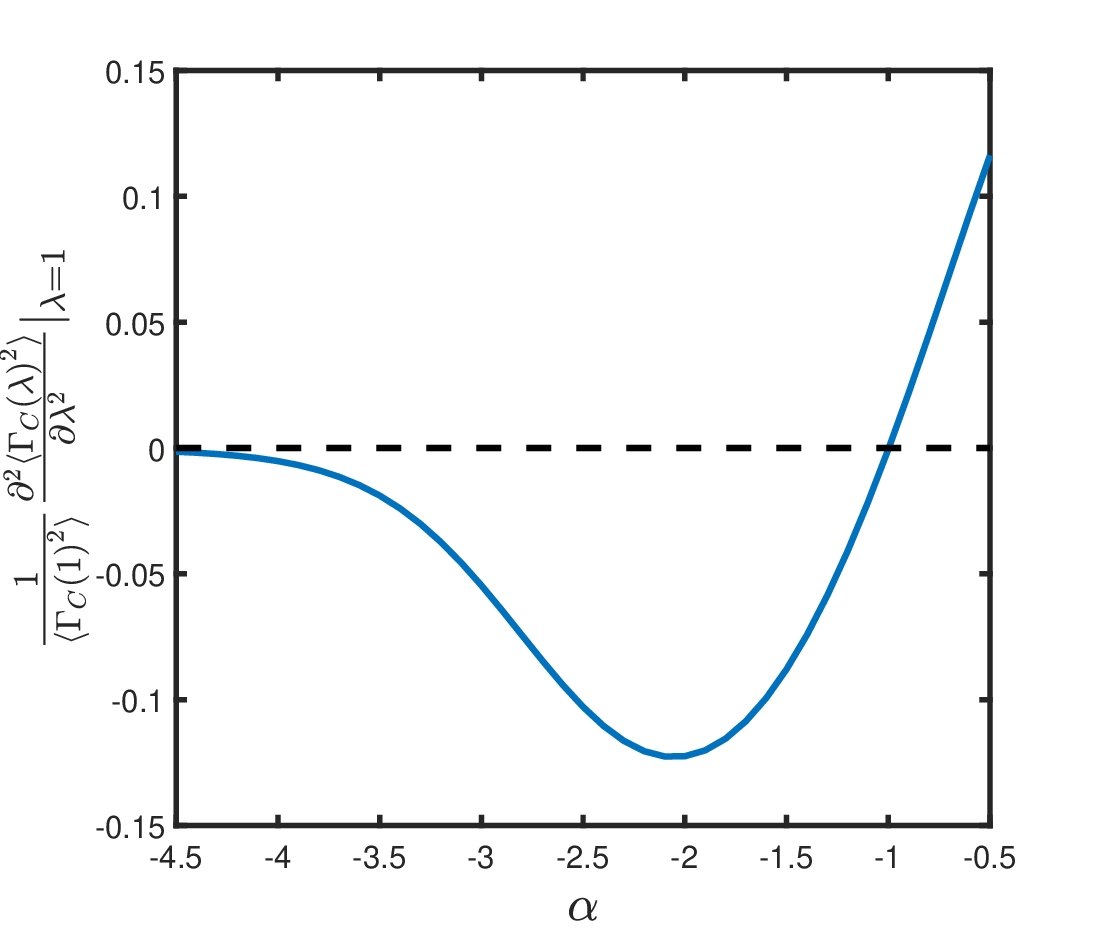}
        \put(0,80){(b)}
    \end{overpic}
    
    \caption{
    Calculation results relate to the second-order moment of circulation in 2-D turbulence with energy spectrum $E(K) \sim K^{\alpha}$.
    (a) The relation between the second-order moment of velocity circulation on a rectangular loop with fixed area and the aspect ratio of the loop, and
    (b) The normalized second derivative of the second moment of circulation when the aspect ratio equals $1$.
    $\lambda$ denotes the aspect ratio of rectangular loops. $\abr{{\Gamma_C (\lambda)}^2}$ is the second-order moment of velocity circulation when the aspect ratio equals $\lambda$.
    }
    \label{2nd moment in inertial ranges}
\end{figure}

Consider an ideal case where the energy or enstrophy inertial range of incompressible 2-D homogeneous isotropic turbulence dominates the contribution of the energy spectrum when calculating the second-order moment of velocity circulation. 
In these cases, the 1-D energy spectrum can be uniformly written in the form of $E(K) \sim K^{\alpha}$, where $K$ is the wavenumber, and $\alpha$ is a scaling exponent with $\alpha = -5/3$ for the energy inertial range and $\alpha = -3$ for the energy inertial range, respectively. Because $E(K) \sim K \hat{E}_{k,l}$ and $K = \sqrt{k^2 + l^2}$, we can set $\hat{E}_{k,l} = E_0 (k^2 + l^2)^{(\alpha - 1)/2}$ ($k^2 + l^2 > 0$) and substitute it into (\ref{circulation second-order moment and energy spectrum}). 
We set $L_x = L_y = L$ and $A/L^2 = 3.6 \times 10^{-5}$, calculate the partial sum of (\ref{circulation second-order moment and energy spectrum}) within the range of $-N \leq k \leq N$ and $-N \leq l \leq N$, where $N = 3 \times 10^4$, then get the result shown in Figure \ref{2nd moment in inertial ranges} (a). 
We also calculated other scaling indices $\alpha$ that depart from inertial ranges. 
An important finding is that only when $\alpha = -1$, the second-order moment of circulation does not vary with the aspect ratio.
Figure \ref{2nd moment in inertial ranges} (b) shows numerically calculated second derivative of the circulation second-order moment with respect to the aspect ratio for different $\alpha$ when the aspect ratio is equal to $1$.
Since the second derivative equal to zero is a necessary condition for a function to be constant, when $\alpha \neq -1$, the second-order moment of circulation must vary with the aspect ratio. 
It is worth mentioning that when $\alpha \rightarrow - \infty$, the second derivative with aspect ratio equal to $1$ will tends to zero. 
This is because almost only the energy spectrum around $K=1$ contributes to the second-order moment of circulation, and then $\abr{\Gamma_C^2} \approx 32 \pi^2 E_0 A^2 / L^2$.

For the energy or enstrophy inertial range of 2-D turbulence, the second-order moment of velocity circulation on a rectangular loop varies with the aspect ratio of the loop, and therefore, the area rule does not hold in either case. 
More generally, when $E(K) \sim K^\alpha$, the area rule only holds when $\alpha =-1$. 
Consequently, we do not focus on studying the area rule in the inertial ranges.

\section{Derivation of the area rule from the loop equation of 2-D instability-driven turbulence}
\label{Sec_derivation}

\subsection{Probability density function of velocity circulation and area derivative}

Introducing the ensemble average $\abr{\cdot}$, we obtain the PDF of velocity circulation on an orientable loop $C$ fixed in space and time 
\begin{equation}\label{P_def}
    P[C,\Gamma] = \abr{\delta(\Gamma - \Gamma_C)} = \abr{\delta \rbr{\Gamma - \oint _C \boldsymbol{v} \cdot d\boldsymbol{r}}} ,
\end{equation}
where $\delta(x)$ is the Dirac delta function.

Following \citet{Migdal1995}, we introduce the area derivative, particularly for the probability density function (PDF) of velocity circulation.
Consider a functional $U[C,\lambda] = U(\Lambda_{\boldsymbol{G}}[C],\lambda)$, where $C$ is an orientable loop in a 2-D plane, $\lambda$ is a variable independent of $C$, e.g., circulation $\Gamma$, $\Lambda_{\boldsymbol{G}}[C] \equiv \oint_C{\boldsymbol{G} \cdot d\boldsymbol{r}} = \iint_S{(\partial_x G_y - \partial_y G_x) d\sigma}$, and $\boldsymbol{G} = G_x \boldsymbol{e}_x + G_y \boldsymbol{e}_y$ is a global smooth vector field without singular point. 
Considering an area element, $\delta \sigma (\boldsymbol{r})$,  located at $\boldsymbol{r}$ with $\delta S$ is its area and $\delta C$ is its boundary, which is a small closed loop. 
For any orientable $C$ and $\delta C$, we can always find a curve $l$ connecting them to form a new orientable loop $(C+\delta C)$. Letting the endpoint of curve $l$ on $C$ be $L_1$ and the endpoint on $\delta C$ be $L_2$, we can define 
\begin{equation}
    \Lambda_{\boldsymbol{G}}[C+\delta C] \equiv \oint_{C + \delta C}{\boldsymbol{G} \cdot d\boldsymbol{r}} = \oint_{C}{\boldsymbol{G} \cdot d\boldsymbol{r}} + \oint_{\delta C}{\boldsymbol{G} \cdot d\boldsymbol{r}} + \int_{L_1}^{L_2}{\boldsymbol{G} \cdot d\boldsymbol{r}} + \int_{L_2}^{L_1}{\boldsymbol{G} \cdot d\boldsymbol{r}} .
\end{equation}
Because the last two integrals in the RHS of the above equation are both integrals on the curve $l$ with opposite directions, the sum of these two integrals is equal to zero. 
Thus,
\begin{equation}
    \Lambda_{\boldsymbol{G}}[C+\delta C] = \oint_{C}{\boldsymbol{G} \cdot d\boldsymbol{r}} + \oint_{\delta C}{\boldsymbol{G} \cdot d\boldsymbol{r}} .
    \label{Lambda_G C+dC}
\end{equation}
When $\delta S \rightarrow 0$, the second term in the RHS of (\ref{Lambda_G C+dC}) tends to zero, so $\Lambda_{\boldsymbol{G}}[C + \delta C] \rightarrow \Lambda_{\boldsymbol{G}}[C]$ and $U[C + \delta C] \rightarrow U[C]$. Then, the area derivative is defined as
\begin{equation}\label{Definition of area derivative}
    \frac{\delta U}{\delta \sigma (\boldsymbol{r})} = \lim_{\delta S \rightarrow 0} \frac{U[C+\delta C,\lambda] - U[C,\lambda]}{\delta S} .
\end{equation}

Based on (\ref{Definition of area derivative}), we can calculate the area derivative of functional $P(C,\Gamma)$:
\begin{equation}
    P[C+\delta C,\Gamma] - P[C,\Gamma] = \abr{\delta \rbr{\Gamma - \oint _{C+\delta C} \boldsymbol{v} \cdot d\boldsymbol{r}} - \delta \rbr{\Gamma - \oint _C \boldsymbol{v} \cdot d\boldsymbol{r}}} ,
\end{equation}
where
\begin{equation}
    \oint _{C+\delta C} \boldsymbol{v} \cdot d\boldsymbol{r} = \oint _C \boldsymbol{v} \cdot d\boldsymbol{r} + \oint _{\delta C} \boldsymbol{v} \cdot d\boldsymbol{r} .
\end{equation}
Because the velocity and its derivative in fluid are finite, when $\delta S \rightarrow 0$, the velocity circulation on loop $\delta C$ also tends towards zero. And then (cf. \S \ref{Sec_Appendix_A}),
\begin{equation} \label{2.6}
    P[C+\delta C,\Gamma] - P[C,\Gamma] = \abr{\delta ' \rbr{\Gamma - \oint _C \boldsymbol{v} \cdot d\boldsymbol{r}} \rbr{-\oint _{\delta C} \boldsymbol{v} \cdot d\boldsymbol{r}}} + o\rbr{\oint _{\delta C} \boldsymbol{v} \cdot d\boldsymbol{r}} .
\end{equation}
Applying the Stokes theorem to the loop integral on $\delta C$, and considering that $\delta S$ is very small and is independent of time, the above equation can be expressed as
\begin{equation}
    P[C+\delta C,\Gamma] - P[C,\Gamma] = -\delta S \abr{\delta ' \rbr{\Gamma - \oint _C \boldsymbol{v} \cdot d\boldsymbol{r}} \omega(\boldsymbol{r},t)} + o(\omega (\boldsymbol{r},t) \delta S) ,
\end{equation}
where $\omega(\boldsymbol{r},t)$ is the vorticity evaluated at the location of the area element surrounded by $\delta C$.
From (\ref{Definition of area derivative}), we obtain
\begin{equation}\label{first area derivative of PDF}
    \frac{\delta P}{\delta \sigma (\boldsymbol{r})} = -\abr{\delta ' \rbr{\Gamma - \oint _C \boldsymbol{v} \cdot d\boldsymbol{r}} \omega(\boldsymbol{r},t)} .
\end{equation}
Then, the second-order area derivative is calculated by taking the first-order area derivative to (\ref{first area derivative of PDF}):
\begin{equation}\label{second area derivative of PDF}
    \frac{\delta^2 P}{\delta \sigma (\boldsymbol{r}_1) \delta \sigma (\boldsymbol{r}_2)} = \abr{\delta '' \rbr{\Gamma - \oint _C \boldsymbol{v} \cdot d\boldsymbol{r}} \omega(\boldsymbol{r}_1,t) \omega(\boldsymbol{r}_2,t)} ,
\end{equation}
where $\boldsymbol{r}_1$ and $\boldsymbol{r}_2$ are two position vectors.

\subsection{Loop equation} \label{section: Loop equation}

We consider 2-D incompressible instability-driven flow with the momentum equation
\begin{equation}\label{general momentum equation}
    \frac{\partial \boldsymbol{v}}{\partial t} + \boldsymbol{v} \cdot \nabla \boldsymbol{v} = - {\frac{1}{\rho} \nabla p} + \mathscr{L} \boldsymbol{v} ,
\end{equation}
where the linear operator
\begin{equation}\label{linear operator L}
    \mathscr{L} = -\alpha - \sum_{k=1}^{N} \beta_{k} (-\nabla^2)^{k} .
\end{equation}
$\mathscr{L} \boldsymbol{v}$ includes linear damping and  (hyper-)viscous effects. 
By designing the parameters $\alpha$ and $\beta_k$, we can prescribe wavenumbers with linear growth or damping rates to describe an instability-driven flow \citep{Mickelin2018,Linkmann2019,Linkmann2020,vanKan2022,vanKan2023}.

Performing loop integration to (\ref{general momentum equation}), moving the nonlinear term to the RHS, and considering that under incompressible conditions, the pressure gradient term will disappear due to loop integration, we obtain
\begin{equation}\label{loop integral of v/t}
    \oint _C \frac{\partial \boldsymbol{v}}{\partial t} \cdot d\boldsymbol{r} = -\oint _C (-v \omega dx + u \omega dy) + \oint _C (\mathscr{L}u dx + \mathscr{L}v dy).
\end{equation}
According to the Biot-Savart theorem, for incompressible flow, velocity can be expressed as
\begin{subequations}\label{2D Biot-Savart}
    \begin{align}
    u(\boldsymbol{x},t) &= \frac{1}{2 \pi} \iint \frac{\eta}{|\boldsymbol{\xi}|^2} \omega(\boldsymbol{x} + \boldsymbol{\xi},t) d\xi d\eta , \\
    v(\boldsymbol{x},t) &= -\frac{1}{2 \pi} \iint \frac{\xi}{|\boldsymbol{\xi}|^2} \omega(\boldsymbol{x} + \boldsymbol{\xi},t) d\xi d\eta ,
    \end{align}
\end{subequations}
where $\boldsymbol{x} = (x,y)$, $\boldsymbol{\xi} = (\xi,\eta)$, $|\boldsymbol{\xi}|^2 = \xi ^2 + \eta ^2 $.

Take partial derivatives of (\ref{P_def}) with respect to time $t$ and circulation $\Gamma$, we obtain
\begin{equation}\label{p_tg}
    \frac{\partial}{\partial \Gamma} \frac{\partial}{\partial t} P(C,\Gamma) = \abr{\delta '' \rbr{\Gamma - \oint _C \boldsymbol{v} \cdot d\boldsymbol{r}} \rbr{- \oint _C \frac{\partial \boldsymbol{v}}{\partial t} \cdot d\boldsymbol{r}}} .
\end{equation}
Substituting (\ref{loop integral of v/t}) and (\ref{2D Biot-Savart}), and under the assumption that the order of averaging, loop integration, and area integration can be exchanged, then considering (\ref{first area derivative of PDF}) and (\ref{second area derivative of PDF}), we obtain the loop equation \citep{Migdal2019}
\begin{equation} \label{Loop Eq, two terms}
    \begin{aligned}
        \frac{\partial}{\partial \Gamma} \frac{\partial}{\partial t} P(C,\Gamma) = & \underbrace{
        \frac{1}{2 \pi} \oint _C \sbr{\rbr{dy {\iint \frac{\eta}{|\boldsymbol{\xi}|^2} d\xi d\eta} + dx {\iint \frac{\xi}{|\boldsymbol{\xi}|^2} d\xi d\eta}} \frac{\delta^2 P}{\delta \sigma (\boldsymbol{x}) \delta \sigma (\boldsymbol{x+\xi})}}
        }_{H}
        \\
        + & \underbrace{
        \frac{1}{2 \pi} \oint_{C} {\sbr{\rbr{dy \iint{\frac{\xi}{|\boldsymbol{\xi}|^2} d\xi d\eta} - dx \iint{\frac{\eta}{|\boldsymbol{\xi}|^2} d\xi d\eta}} \mathscr{L} \frac{\partial}{\partial \Gamma} \frac{\delta P(C,\Gamma)}{\delta \sigma (\boldsymbol{x} + \boldsymbol{\xi})}}}
        }_{I} ,
    \end{aligned}
\end{equation}
where $H$ and $I$ denote the contributions from nonlinear and linear terms, respectively.

\subsection{Derivation of area rule}

Inspired by the area rule proposed by \citep{Migdal2019}, and under the assumption of homogeneity, we speculate that the loop equation (\ref{Loop Eq, two terms}) has a solution
\begin{equation}\label{area rule}
    P(C,\Gamma) = P (A_C,\Gamma) ,
\end{equation}
where
\begin{equation}
\label{definition of area enclosed by loop}
    A_C = \iint_{S_C}{d\sigma} = \frac{1}{2} \oint_C {xdy-ydx}
\end{equation}
is the area enclosed by the orientable loop $C$.

For fully developed turbulence in a statistically steady state, the left-hand side of the loop equation (\ref{Loop Eq, two terms}) is zero, then we need to prove that (\ref{area rule}) can make
\begin{equation}\label{formula to be proven (area rule)}
    H + I = 0.
\end{equation}
In the following subsections \S \ref{sec_H (area rule)} and \S \ref{sec_I (area rule)}, we prove that $H$ and $I$ are both equal to zero.

\subsubsection{Proof of $H=0$} \label{sec_H (area rule)}

Considering $\frac{\delta P}{\delta \sigma (\boldsymbol{x})}$, which is the first-order area derivative of PDF $P(C,\Gamma)$. With expression (\ref{area rule}), we obtain
\begin{equation}
    \frac{\delta P(C,\Gamma)}{\delta \sigma (\boldsymbol{x})} = \frac{\partial P}{\partial A_C} \frac{\delta A_C}{\delta \sigma (\boldsymbol{x})} .
\end{equation}
Note that $\frac{\delta A_C}{\delta \sigma (\boldsymbol{x})} = 1$, so
\begin{equation}\label{firsr-order area derivative of P (area rule)}
    \frac{\delta P(C,\Gamma)}{\delta \sigma (\boldsymbol{x})} = \frac{\partial P}{\partial A_C} .
\end{equation}

Then calculate the second-order area derivative $\frac{\delta^2 P(C,\Gamma)}{\delta \sigma (\boldsymbol{x}_1) \delta \sigma (\boldsymbol{x}_2)}$. Taking the area derivative of (\ref{firsr-order area derivative of P (area rule)}), we get
\begin{equation}\label{second-order area derivative of P (area rule)}
\frac{\delta^2 P(C,\Gamma)}{\delta \sigma (\boldsymbol{x}_1) \delta \sigma (\boldsymbol{x}_2)} = \frac{\partial^2 P}{{\partial A_C}^2} .
\end{equation}

Therefore, $H=0$ is equivalent to 
\begin{equation}\label{H, area rule}
    \frac{1}{2 \pi} \frac{\partial^2 P}{{\partial A_C}^2} \oint_C \rbr{dy \iint{\frac{\eta}{|\boldsymbol{\xi}|^2} d\xi d\eta} + dx \iint{\frac{\xi}{|\boldsymbol{\xi}|^2} d\xi d\eta}} = 0 .
\end{equation}

Although the surface integral is not convergent, considering that the integrand functions in the two surface integrals are odd functions with respect to $\eta$ and $\xi$, respectively. In the sense of taking Cauchy's principal value, we can consider it equal to zero, so $H=0$.

\subsubsection{Proof of $I=0$} \label{sec_I (area rule)}

Secondly, \textcolor{blue}{we} prove that (\ref{area rule}) makes $I=0$. According to (\ref{firsr-order area derivative of P (area rule)}) and (\ref{Loop Eq, two terms}), we can obtain
\begin{equation}
    I = \frac{1}{2 \pi} \oint_{C} {\sbr{\rbr{dy \iint{\frac{\xi}{|\boldsymbol{\xi}|^2} d\xi d\eta} - dx \iint{\frac{\eta}{|\boldsymbol{\xi}|^2} d\xi d\eta}} \mathscr{L} \frac{\partial}{\partial \Gamma} \frac{\partial P}{\partial A_C}}} .
\end{equation}
Because $\mathscr{L} \frac{\partial}{\partial \Gamma} \frac{\partial P}{\partial A_C}$ is independent of $\boldsymbol{\xi}$, we can extract it beyond the two surface integrals. Then, following the previous proof of $H=0$, we can obtain $I=0$.

Note that this section's derivation only shows that the area rule of velocity circulation is a potential solution to the loop equation (\ref{Loop Eq, two terms}) and Migdal's derivation can be generalized to scales beyond the inertial range, at least for the special instability-driven 2-D system (\ref{general momentum equation}) and (\ref{linear operator L}). 
However, as we showed in \S \ref{Sec_2ndmoment}, additional conditions, such as the spectral exponent $-1$, are required for the existence of the area rule.

\section{Numerical simulation}
\label{Sec_numerics}

We numerically studied 2-D turbulence driven by negative viscosity and dissipated by hyperviscosity and linear damping,
\begin{equation}
\label{governing_eq}
    \frac{\partial \omega}{\partial t} + \boldsymbol{u} \cdot \nabla \omega = -\alpha \omega - \lambda \nabla^2 \omega - \nu \nabla^8 \omega ,
\end{equation}
where $\alpha$, $\lambda$, and $\nu$ are positive constants. In a periodic domain with size $L_0 \times L_0$, we can define two dimensionless parameters $R_l = {\alpha L_0^2}/{\lambda}$ and $R_s = {\nu}/{\lambda L_0^6}$ to measure the dissipation of the system at large and small scales, respectively. 
Specifically, in our numerical simulation, $\alpha = 0.018$, $\lambda = 2 \times 10^{-5}$ and $\nu = 1 \times 10^{-16}$. We perform numerical simulations in a doubly periodic domain of size $2\pi \times 2\pi$ with a resolution of $2048 \times 2048$. We used the fourth-order Runge-Kutta method for the time integration and the pseudospectral method for the spatial integration \citep[cf.][]{Xie_Buhler_2019}. 

\begin{figure}
    \centering    \includegraphics[width=0.5\linewidth]{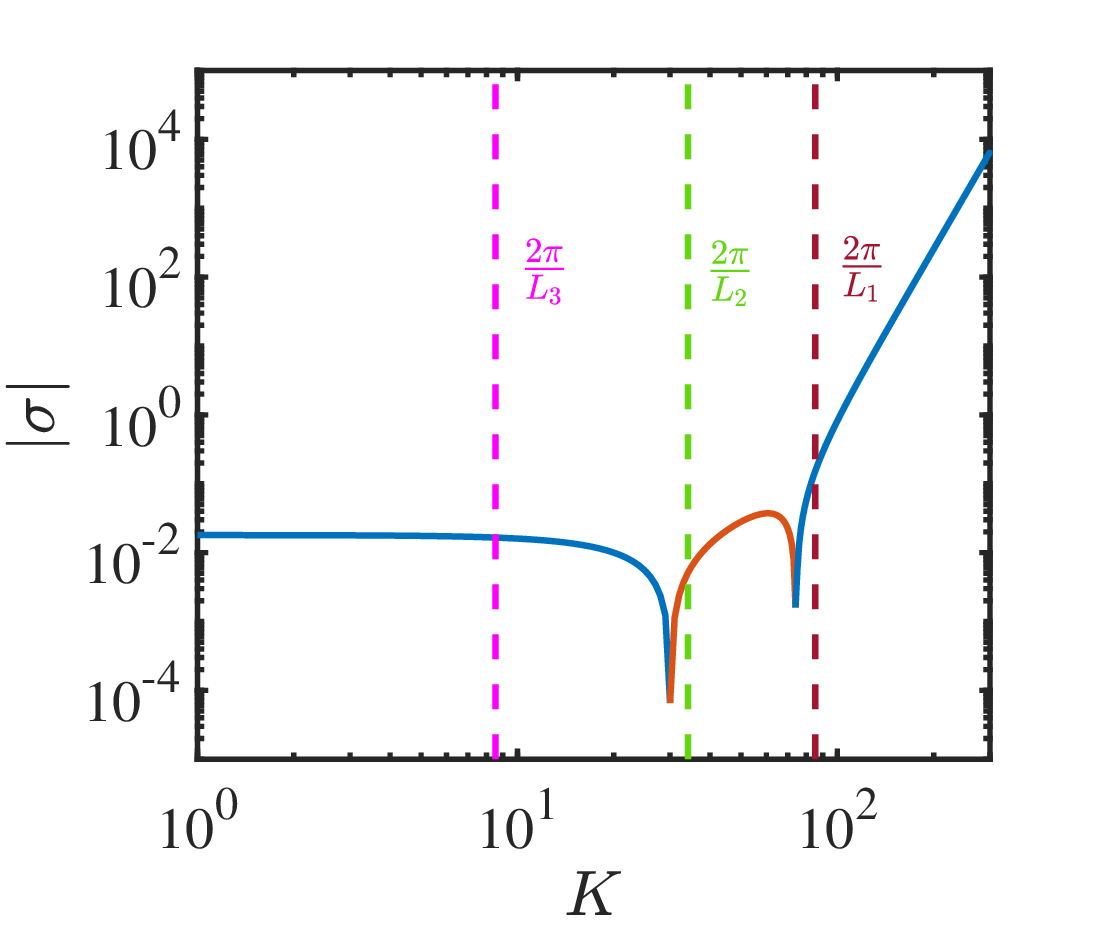}
    \caption{The absolute value of the linear growth rate $\sigma$ in spectral space. $K$ is wavenumber.
    The blue and red curves correspond to $\sigma < 0$ and $\sigma > 0$, respectively.
    The three vertical dashed lines denote the loop scales studied in Fig. \ref{simple loops PDF}.}
    \label{L hat}
\end{figure}

Considering a Fourier mode $e^{i(kx+ly)+\sigma t}$, the linear growth rate of (\ref{governing_eq}) is
\begin{equation}
    \sigma = -\alpha + \lambda K^2 - \nu K^8,
\end{equation}
where $K = \sqrt{k^2 + l^2}$ is the modulus of the wave vector $\boldsymbol{k} = (k,l)$. At a certain scale, when $\sigma > 0$ energy is injected, and when $\sigma < 0$ energy is dissipated.
Figure \ref{L hat} shows the linear growth rate $\sigma$ in our simulation.
When the wavenumber $K \in \sbr{30,74}$, energy is injected into the flow field, and in other wavenumber ranges, energy dissipates. 
In our numerical simulation, $l_0 \approx 4\Delta x$, where $\Delta x = 2\pi / 2048$ is the grid spacing and the dissipation scale $l_0 = \sbr{\nu^3 / \abr{\eta} }^{1/24}$ with $\abr{\eta}$ the average enstrophy injection rate. 

\begin{figure}
\centering
\includegraphics[width=0.6\linewidth]{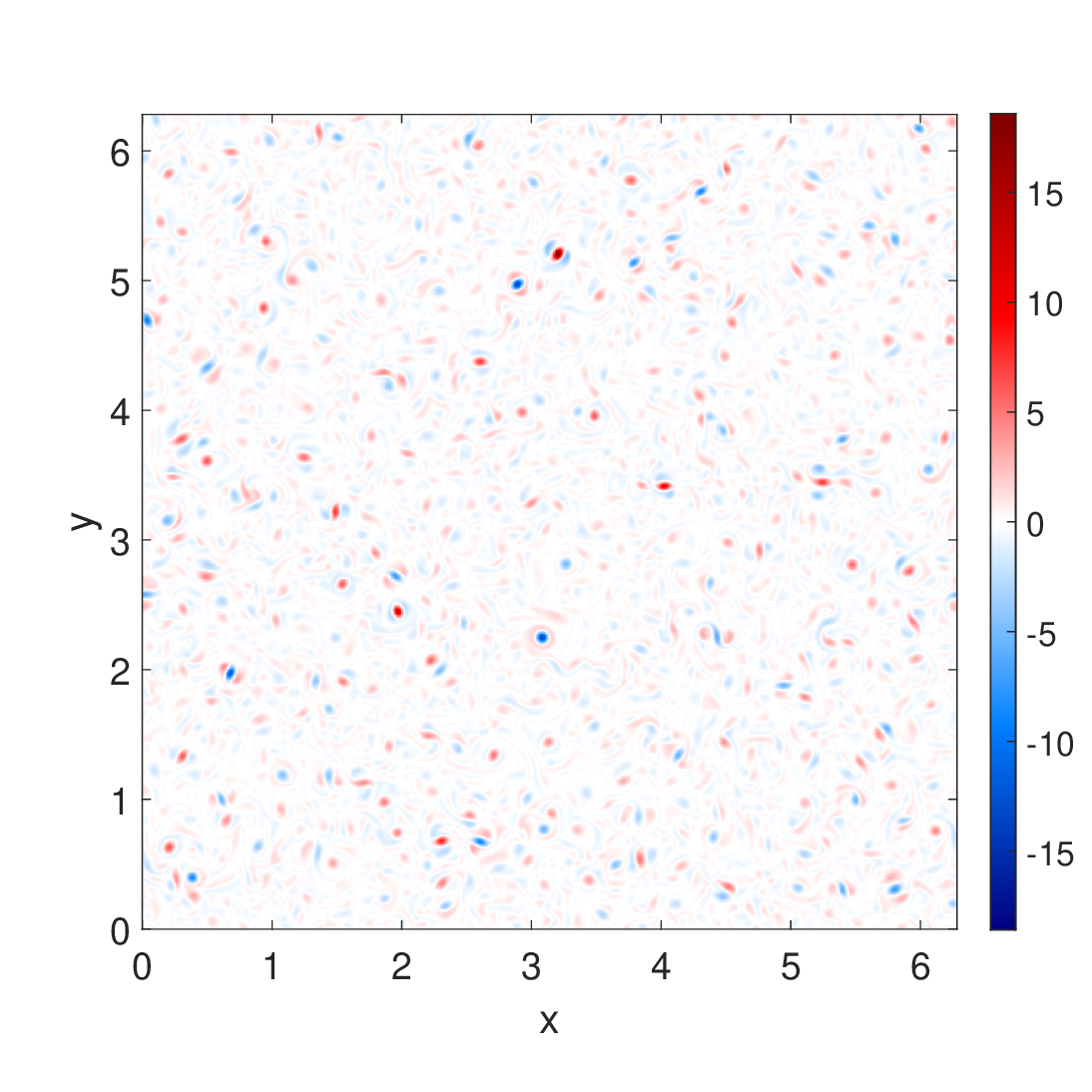}
\caption{A snapshot of the vorticity field at the statistically steady state.}
\label{Figure of vorticity}
\end{figure}

Figure \ref{Figure of vorticity} shows a snapshot of vorticity at the statistically steady state. 
Although there is no nonlinear resistance in our equation, we still observed a sea of shielded vortices, a structure in which the vorticity core is surrounded by a ring of vortices with opposite signs \cite{vanKan2022}. 

\begin{figure}
\centerline{\includegraphics[width=\linewidth]{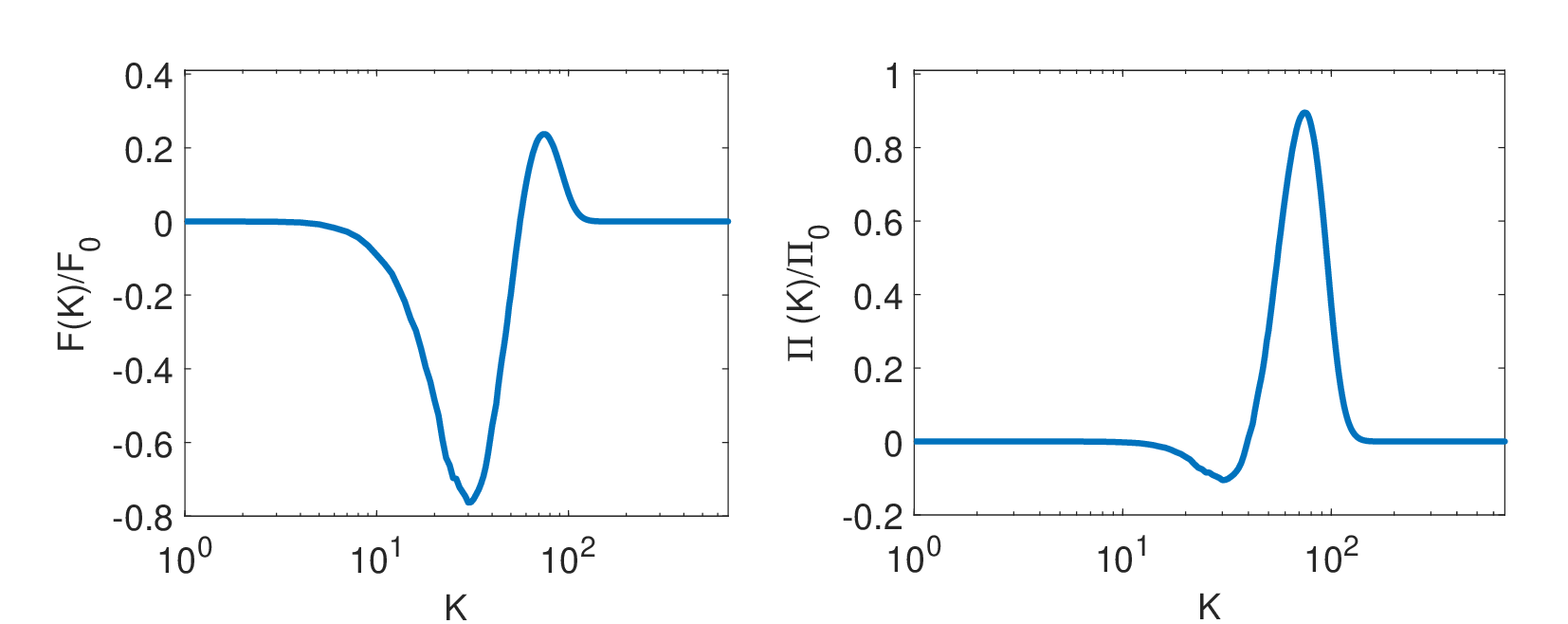}}

\caption{The left figure shows the flux of energy, and the right figure shows the flux of enstrophy. $K$ is wavenumber, $F_0$ is energy injection rate, $\Pi_0$ is enstrophy injection rate.}
\label{Figure of flux}
\end{figure}

\begin{figure}
\centering
\includegraphics[width=0.6\linewidth]{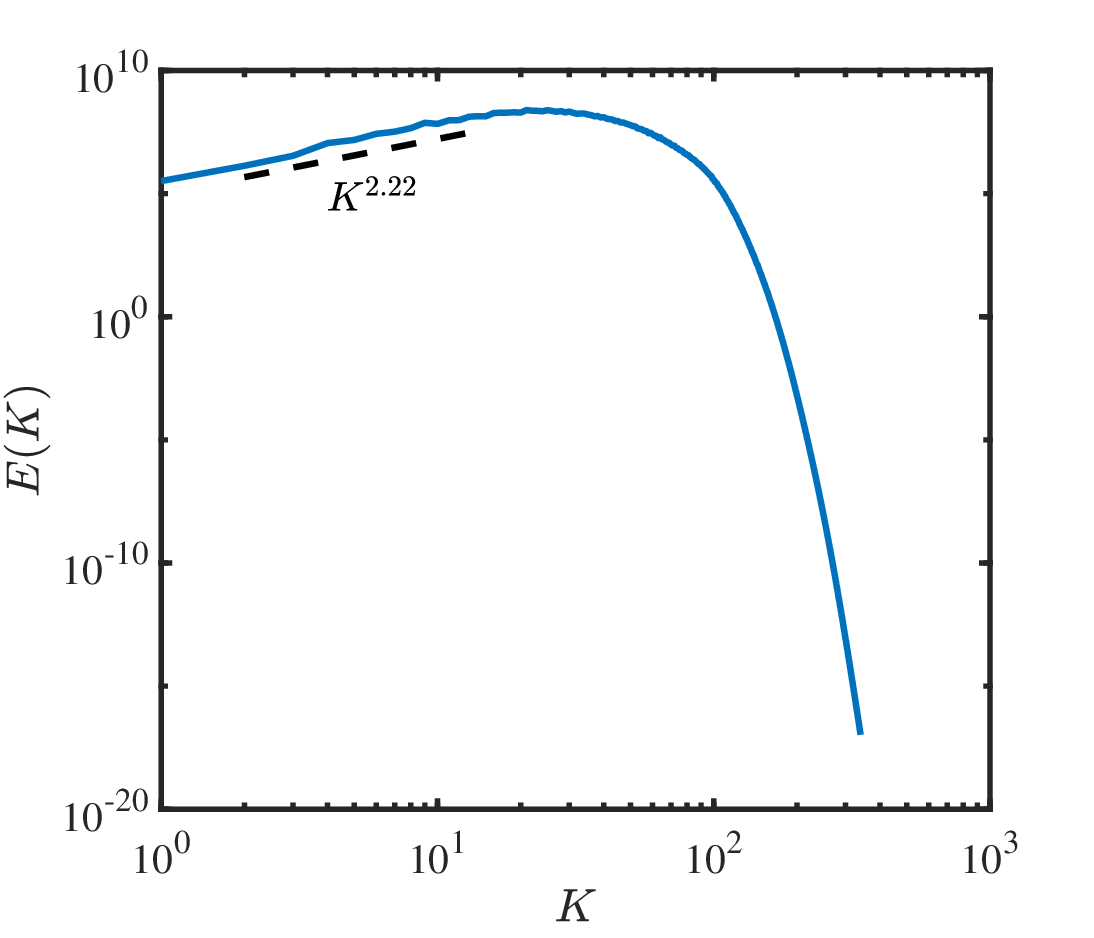}

\caption{The energy spectrum $E(K)$ of flow field.}
\label{Figure of energy spectrum}
\end{figure}

Figure \ref{Figure of flux} shows the flux of energy and enstrophy. 
In this simulation, both energy and enstrophy transfer bidirectionally to large and small scales, and there is no inertial range with constant flux.
Thus, this simulation is suitable for studying statistics beyond the inertial range.
Figure \ref{Figure of energy spectrum} shows the energy spectrum. 
This energy spectrum is similar to that of \cite{vanKan2022}'s flow field solely driven by negative viscosity.

\subsection{Rectangular loops}
\label{Subsec_simple loop}

\begin{figure}
\centering
    \begin{overpic}[width=0.32\textwidth]{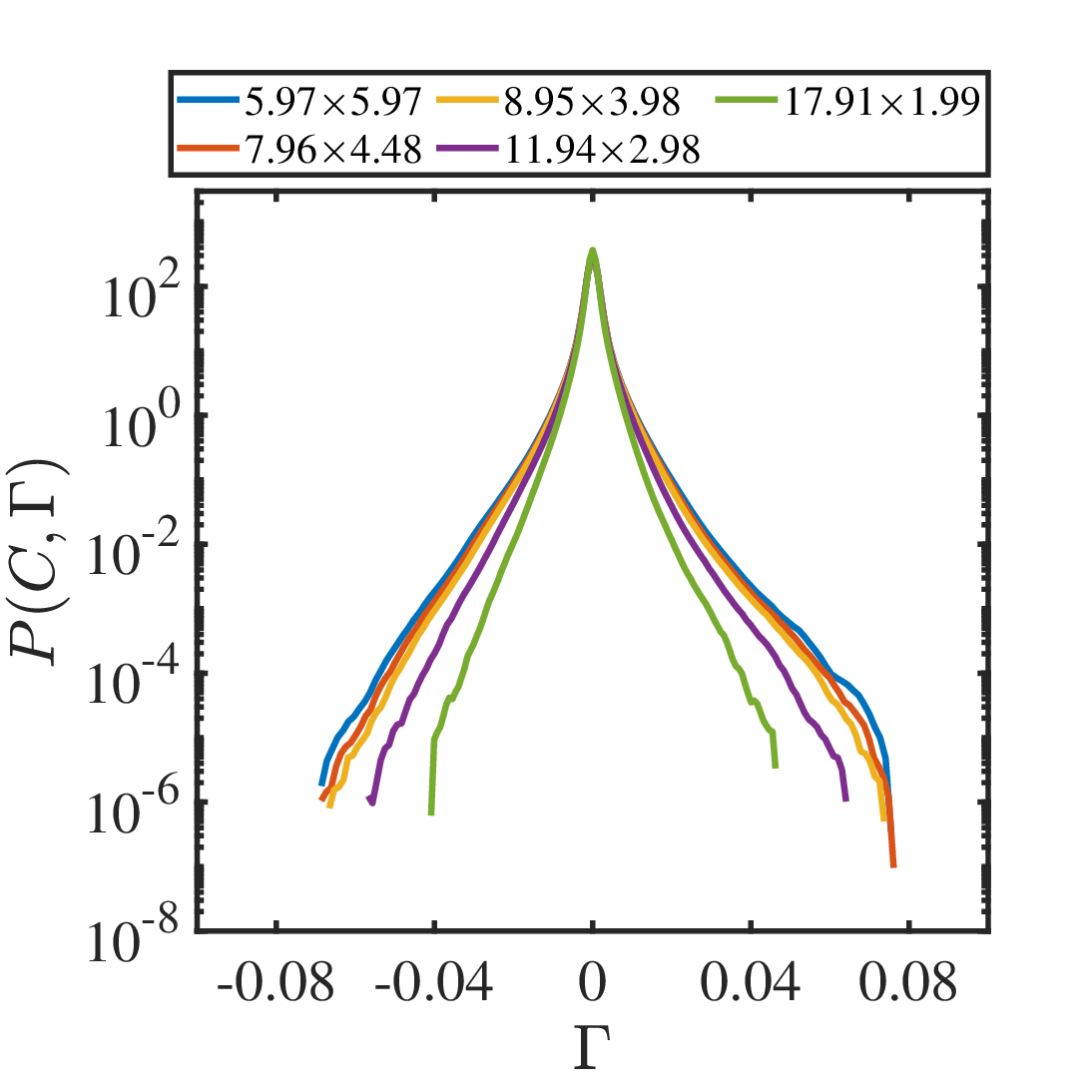}
        \put(0,80){(a)}
    \end{overpic}
    \begin{overpic}[width=0.32\textwidth]{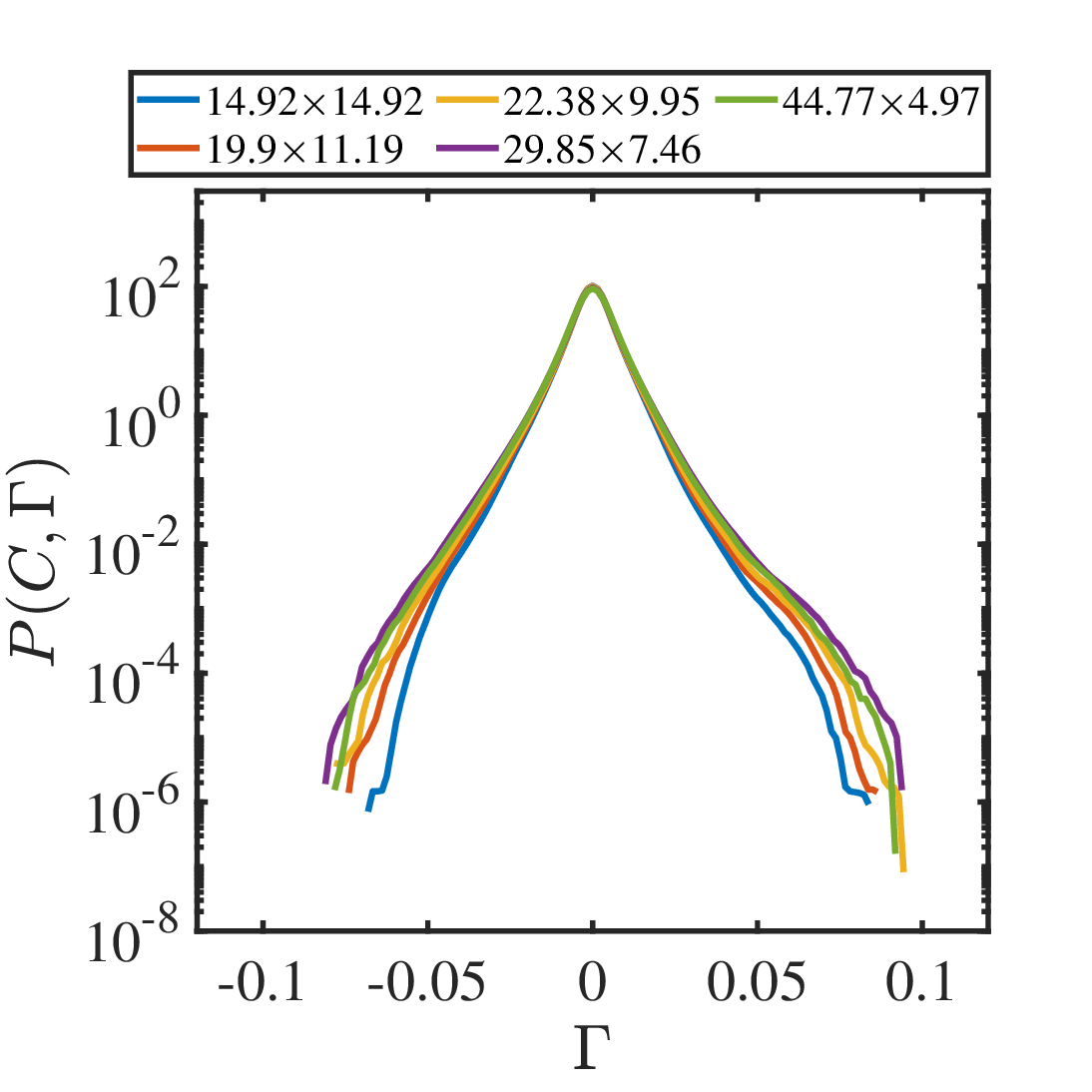}
        \put(0,80){(b)}
    \end{overpic}
    \begin{overpic}[width=0.32\textwidth]{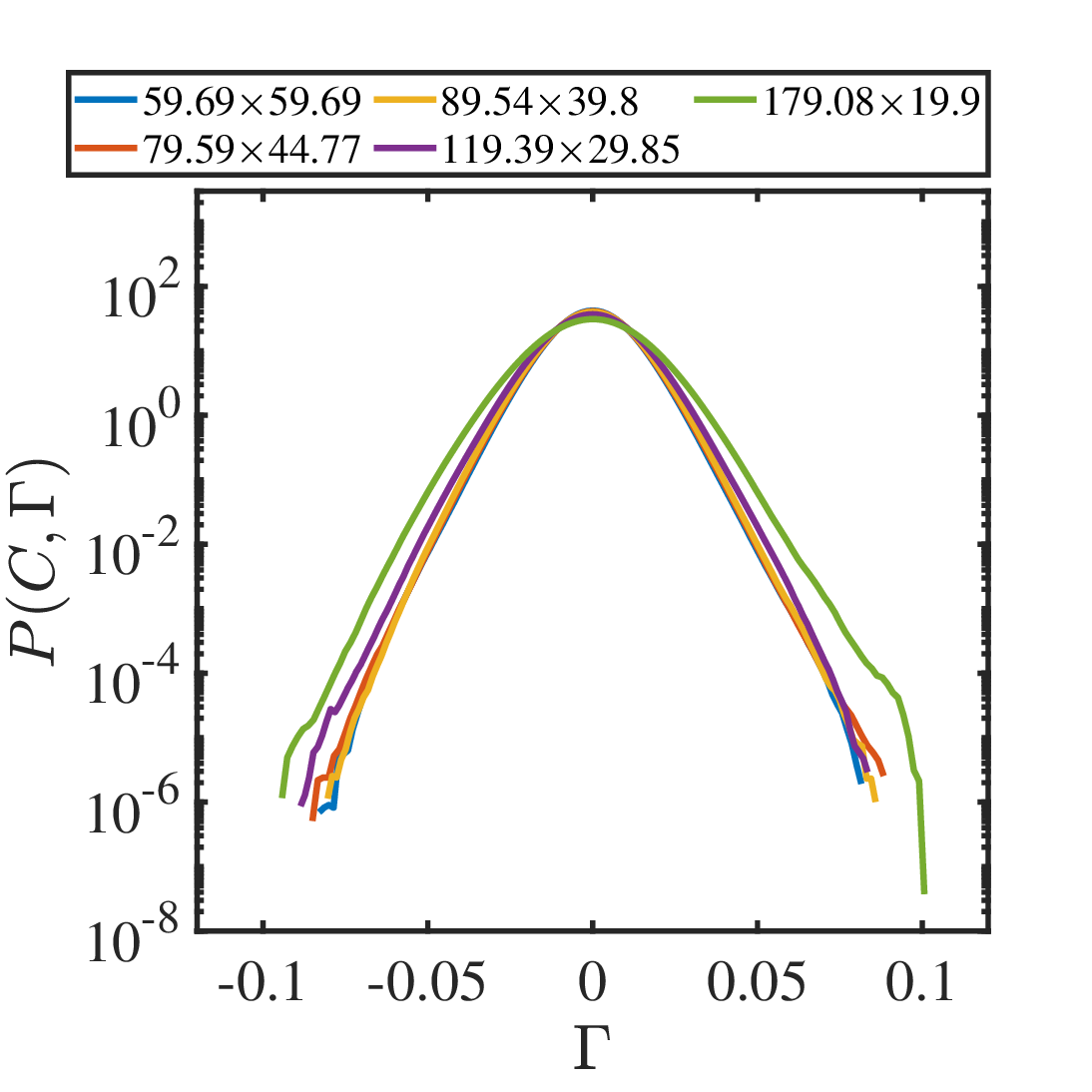}
        \put(0,80){(c)}
    \end{overpic}
    \begin{overpic}[width=0.32\textwidth]{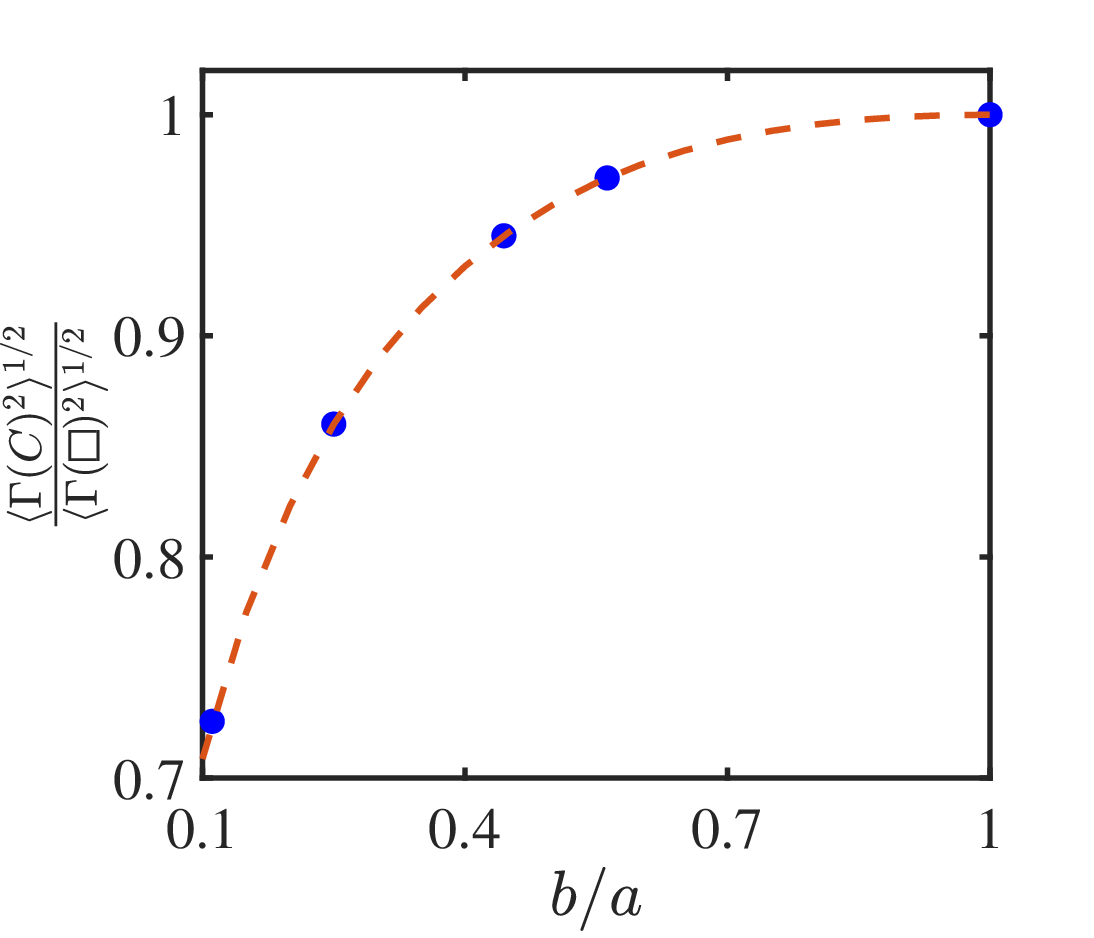}
        \put(0,80){(d)}
    \end{overpic}
    \begin{overpic}[width=0.32\textwidth]{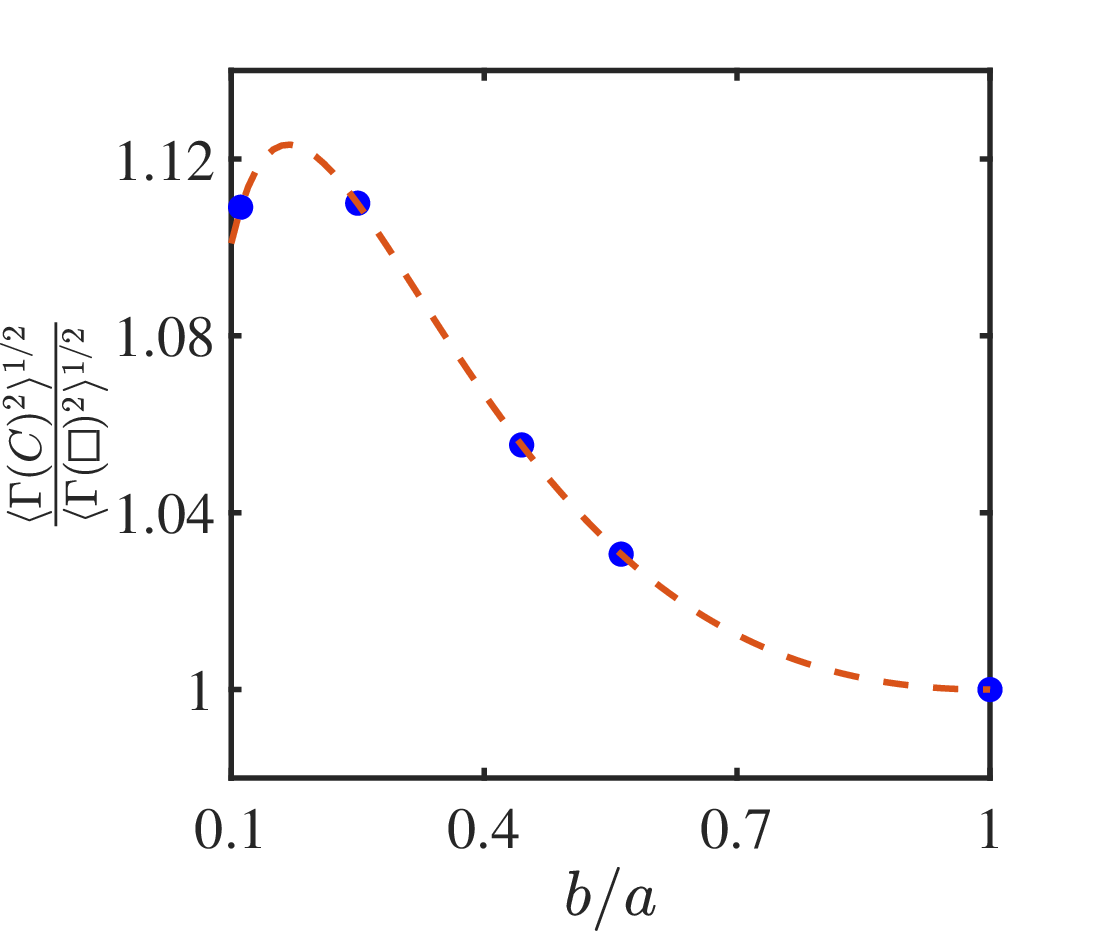}
        \put(0,80){(e)}
    \end{overpic}
    \begin{overpic}[width=0.32\textwidth]{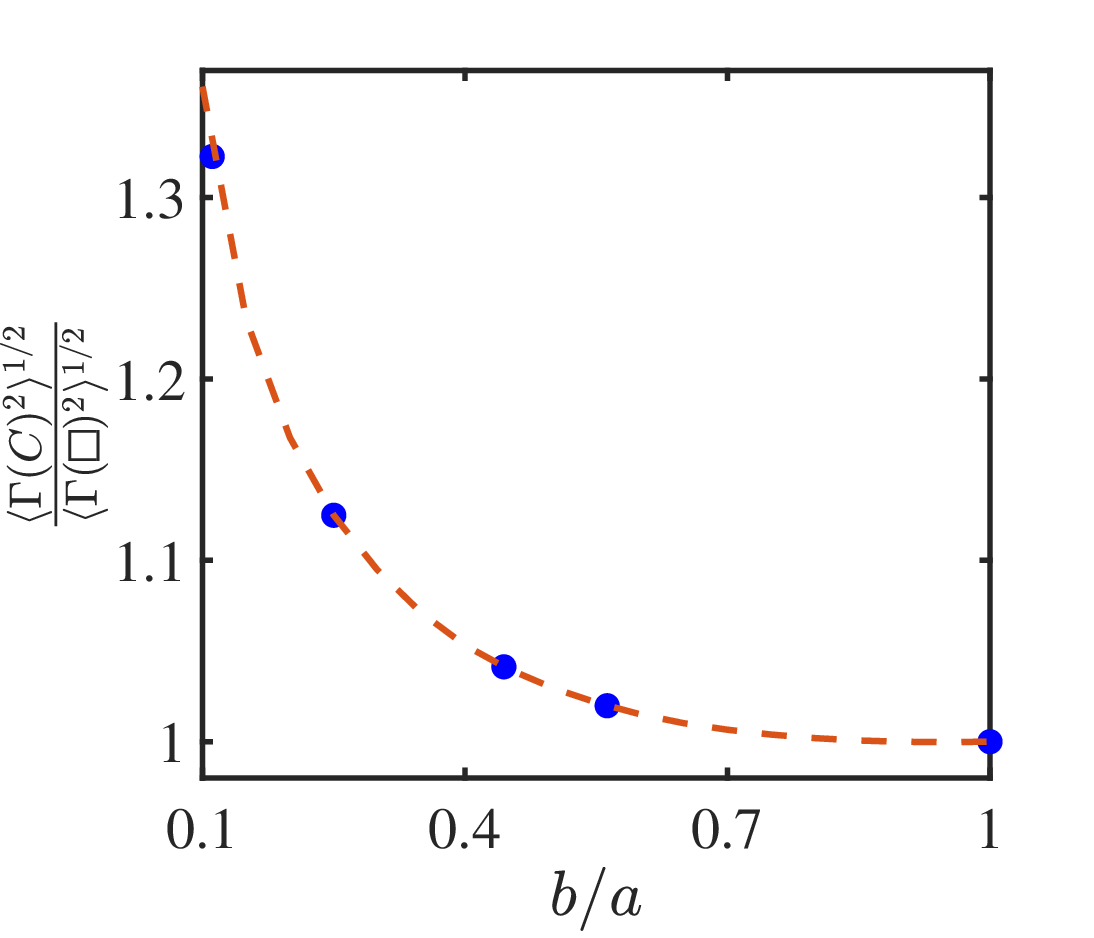}
        \put(0,80){(f)}
    \end{overpic}
    \begin{overpic}[width=0.32\textwidth]{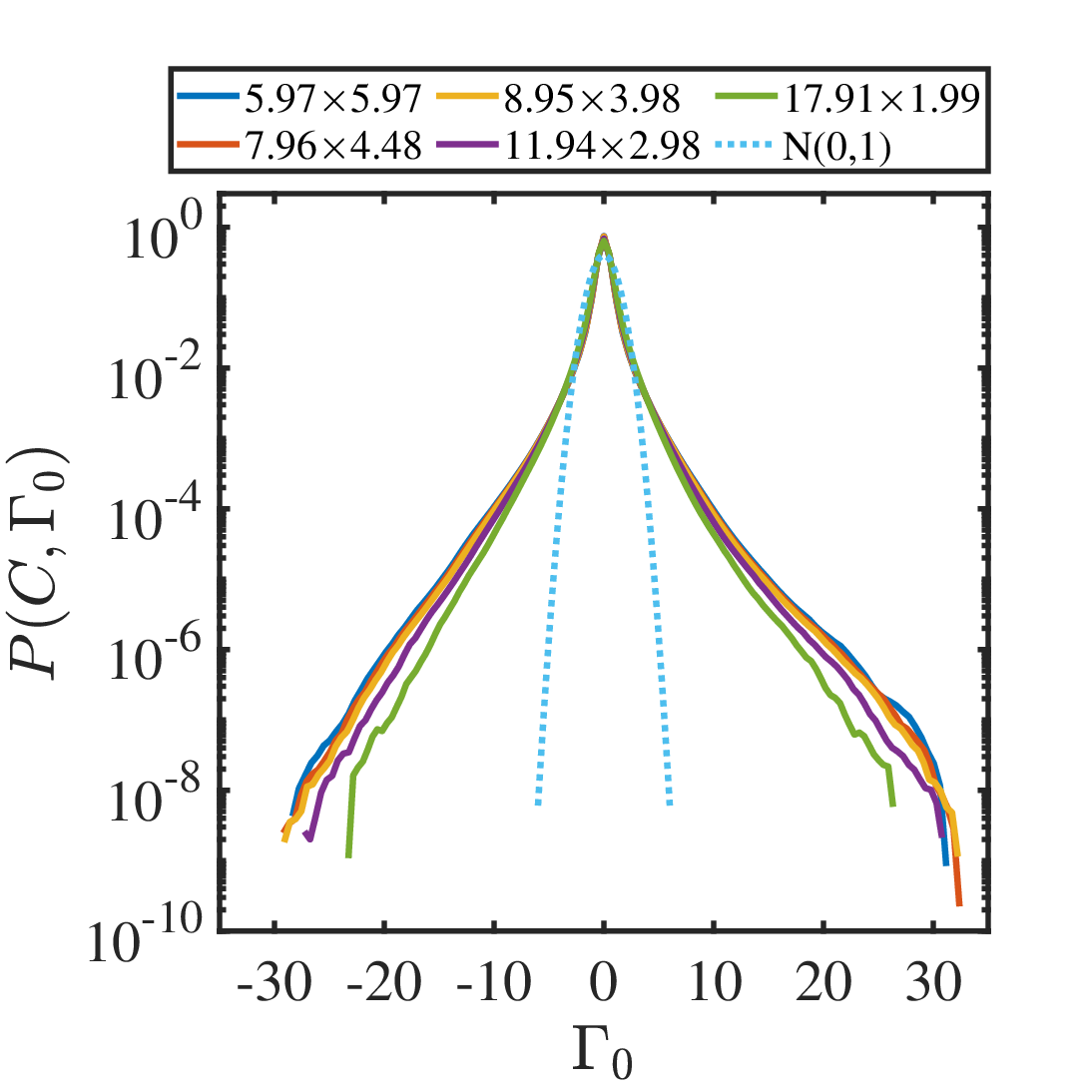}
        \put(0,80){(g)}
    \end{overpic}
    \begin{overpic}[width=0.32\textwidth]{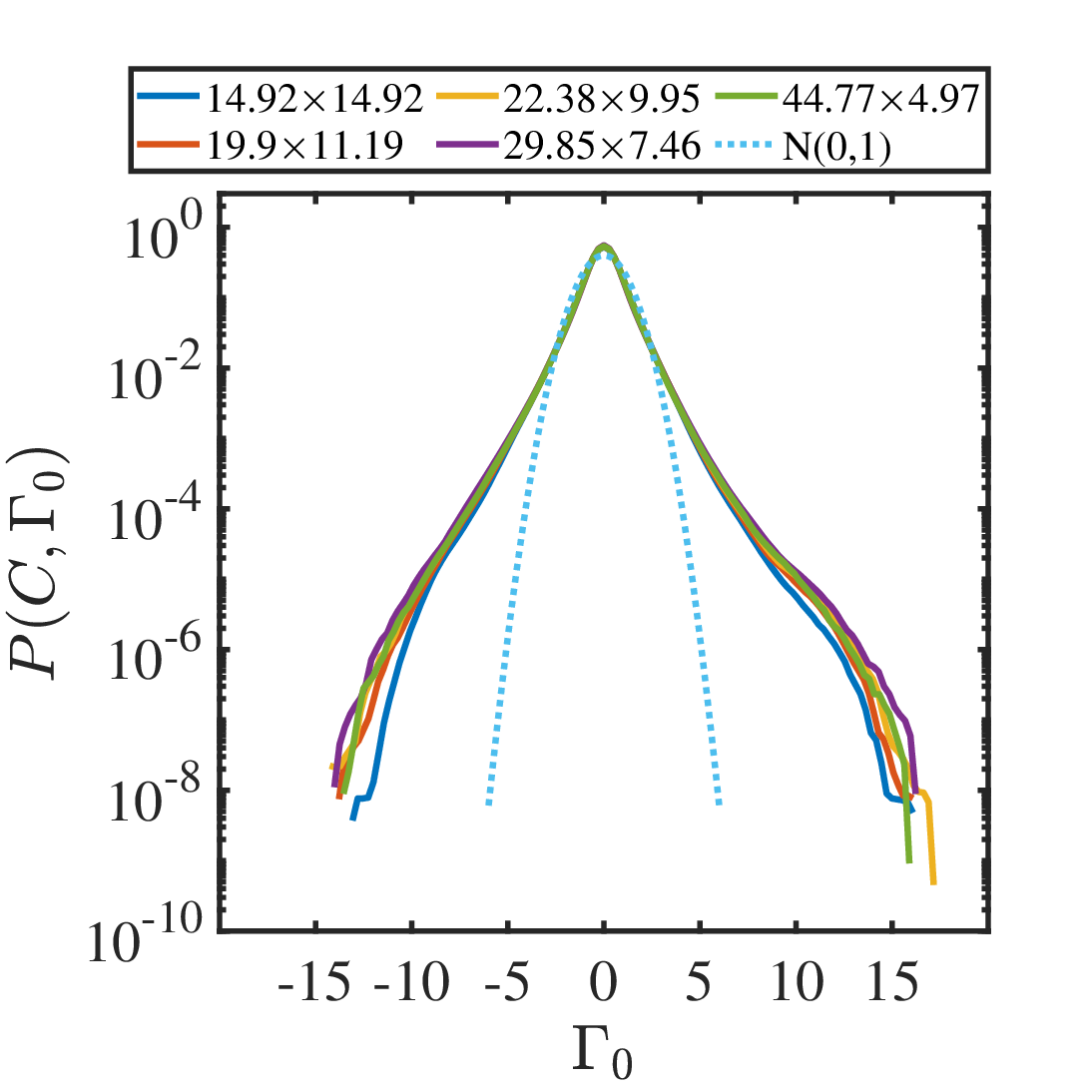}
        \put(0,80){(h)}
    \end{overpic}
    \begin{overpic}[width=0.32\textwidth]{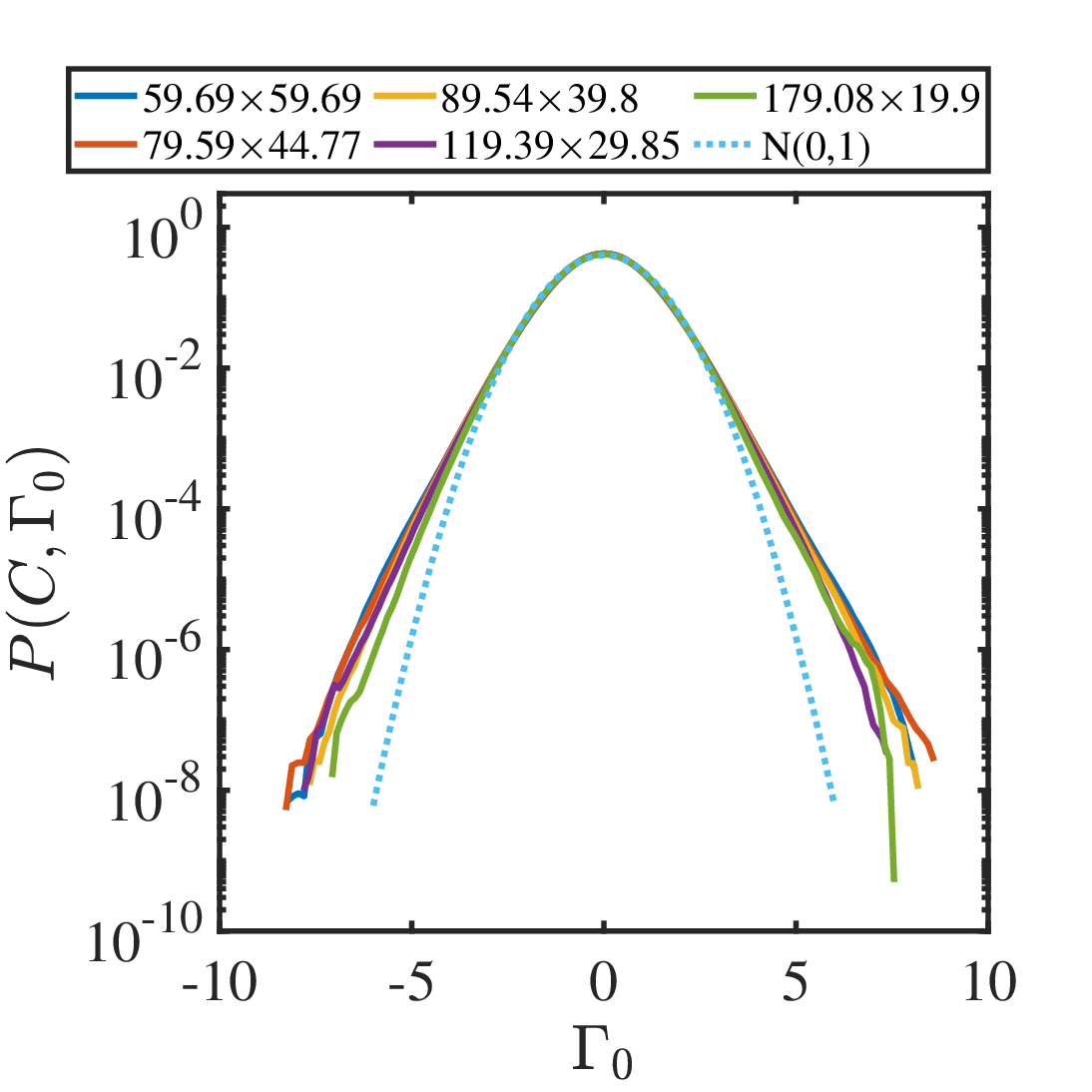}
        \put(0,80){(i)}
    \end{overpic}

\caption{
(a), (b) and (c) show the PDFs corresponding to rectangular loops of sizes $L_1^2 \approx 36 l_0^2$, $L_2^2 \approx 225 l_0^2$ and $L_3^2 \approx 3600 l_0^2$, where $l_0$ is the dissipation scale, with varying aspect ratios.
Here, the wavenumbers corresponding to these scales are marked as vertical dashed lines in Figure \ref{L hat}.
The $a \times b$ in the legend represents the length of the rectangular loop as $a l_0$ and the width as $b l_0$.
Figure (d), (e) and (f) present the dependence of $\abr{\Gamma(C)^2}^{1/2} / {\abr{\Gamma(\Box)^2}^{1/2}}$ on the aspect ratio $b/a$ of the rectangular loops with fixed areas corresponding to (a), (b) and (c), respectively. 
Here, $\abr{\Gamma(C)^2}$ is the second-order moment of circulation on loop $C$ and $\abr{\Gamma(\Box)^2}$ is the second-order moment of circulation on a square loop with the same area. The blue dots show the statistical results of the selected rectangular loops, and the red dashed lines show the calculation results obtained according to formula (\ref{circulation second-order moment and energy spectrum}).
(g), (h) and (i) show the normalized PDFs corresponding to the same loops as (a), (b) and (c), respectively. The definition of $\Gamma_0$ is $\Gamma_0 = \Gamma/\abr{\Gamma(C)^2}^{1/2}$.
}

\label{simple loops PDF}
\end{figure}

Figure \ref{simple loops PDF} shows the variation of the PDF and second-order moment of the circulation on a rectangular loop with varying aspect ratios within the range of energy injection and energy dissipation at both large and small scales. The scale of a loop is defined as $l_C = \sqrt{|A|}$, where $A$ is the area surrounded by the loop defined by (\ref{definition of area enclosed by loop}).
For rectangular loops with the same area but different aspect ratios, Figure \ref{simple loops PDF} (a)-(c) shows that whether their scale is large or small, there are significant differences between the PDF curves, which implies a departure from the area rule. 
Figure \ref{simple loops PDF} (d) shows that when the aspect ratio of a rectangular loop reduces, the second-order moment of velocity circulation decreases for small loops, while for large loops, shown in Figure \ref{simple loops PDF} (f), the second-order moment of velocity circulation increases as the aspect ratio decreases. For medium loops, shown in Figure \ref{simple loops PDF} (e), as the aspect ratio decreases, the second-order moment of velocity circulation first increases and then decreases as the aspect ratio moves away from $1$.

We can learn that the second-order moment of circulation depends on the aspect ratio of rectangular loops through the relationship between the second-order moment of circulation and the energy spectrum of a periodic flow field (\ref{circulation second-order moment and energy spectrum}).
The dashed lines in (d), (e) and (f) of Figure \ref{simple loops PDF} show the normalized second-order moment of velocity circulation variation curves calculated using the energy spectrum through (\ref{circulation second-order moment and energy spectrum}) for different aspect ratios. 
Here, the theoretical results agree well with the numerical results.
We can qualitatively understand the second-order moment's dependence on the aspect ratio: when the area enclosed by the loop is small, the square of kernel's Fourier transform, $|\hat{G}_{k,l}|^2$, spreads in a wide range of wavenumber, so the decreasing part of the energy spectrum contributes significantly to the calculation of the second-order circulation moment, causing the later to decrease with decreasing aspect ratio (Fig. \ref{simple loops PDF} (d)). When the area enclosed by the loop is large, $|\hat{G}_{k,l}|^2$ concentrates in a narrow wavenumber range, therefore almost only the small-wavenumber part of the energy spectrum contribute to the second-order circulation moment, then the later increase as the aspect ratio decreases (Fig. \ref{simple loops PDF} (f)). For a loop with an intermediate enclosing area, the combined impacts of the large- and small-wavenumber ranges of the energy spectrum lead to a non-monotonous behavior of the dependence of the second-order circulation moment on the aspect ratio (Fig. \ref{simple loops PDF} (e)).

Figure \ref{simple loops PDF} (g), (h) and (i) present the PDFs normalized with their respective second-order moments of circulation, and we observe a phenomenon similar to that reported by \cite{Iyer2021}: After normalization, the PDF curves are better collapsed. 
It should be noted that this collapse is not a trivial mathematical result of equating the second-order moment. 
In a semi-logarithmic coordinate system, where the $y$-axis is the logarithmic axis, PDFs normalized by second-order moments may not be closer than the original PDFs. 
For example, $f(x)$ is a standard normal distribution, and $g(x)$ is a PDF that has a tail thicker than $f(x)$. If the second-order moment corresponding to $g(x)$ is less than $1$, after normalizing with the second-order moments, the tail of $g(x)$ becomes thicker and the difference between $g(x)$ and $f(x)$ appears larger.
In addition, unless otherwise specified, the term ``normalized'' mentioned later specifically refers to normalizing using the second-order moment of the velocity circulation.

\begin{figure}
\centering
\includegraphics[height=6cm,width=6.25cm]{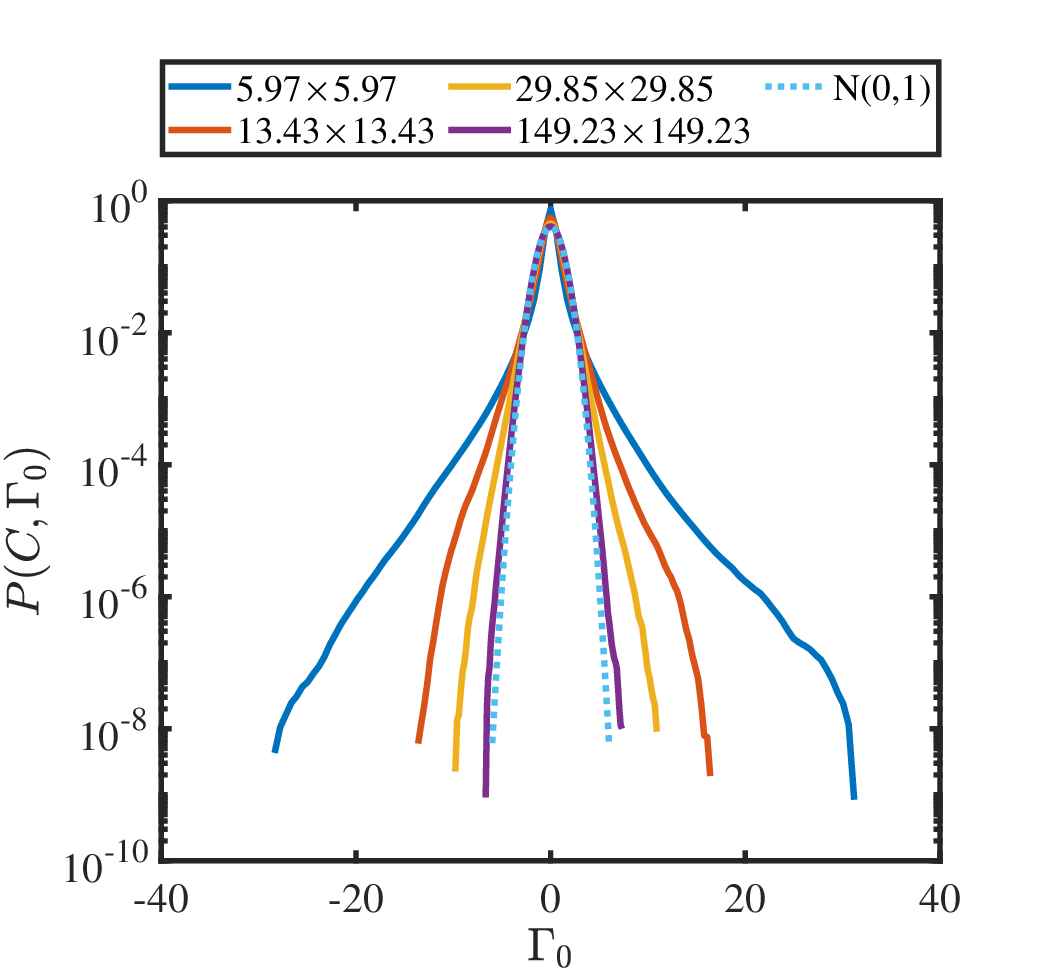}

\caption{Normalized PDFs of square loops with different areas. $\Gamma_0 = \Gamma / \sqrt{s_\Gamma}$ is normalized circulation, where $s_\Gamma$ is the second-order moment of velocity circulation. The $a \times b$ in legend means the curve corresponds to a rectangle with length $a l_0$ and width $b l_0$.}
\label{Normalized PDFs of different areas}
\end{figure}

Considering that the velocity circulation is an averaged (or filtered) vorticity, this section illustrates the similarity between PDFs of velocity circulations and vorticity fluctuations.
Figure \ref{Normalized PDFs of different areas} shows the comparison between normalized PDFs of velocity circulation with different areas.
As the area increases, the statistics become less intermittent and closer to a Gaussian distribution, which resembles the feature of vorticity structure functions in 2D turbulence \citep{Boffetta2002}.

\subsection{8-loops}

To determine whether the statistical behavior of the velocity circulation is determined by the scalar area or vector area of the loop, \cite{Makoto1993} and \cite{Iyer2019} studied the velocity circulation on 8-loops with edges lying in the inertial range of 3-D turbulence. 
They constructed the 8-loop consisting of two squares with side lengths $L_1$ and $L_2$ with fixed $d = L_1-L_2$, and varied the size of $L_1$. 
They numerically obtained the relation between the second-order moment of velocity circulation and the scalar area sum of 8-loops. 
Based on the K41 theory, they concluded that the statistical behavior of velocity circulation depends on the scalar area of the loop. 
Considering that 2-D turbulence and 3-D turbulence have different energy transfer directions, and K41 theory fails when the inertial range does not exist, we cannot guarantee that the conclusions given by Umeki and Iyer et al. still hold in the 2-D instability-driven turbulence. 
Therefore, we also study the normalized statistics of circulations over 8-loops.

\begin{figure}
\centerline{\includegraphics[width=0.4\textwidth]{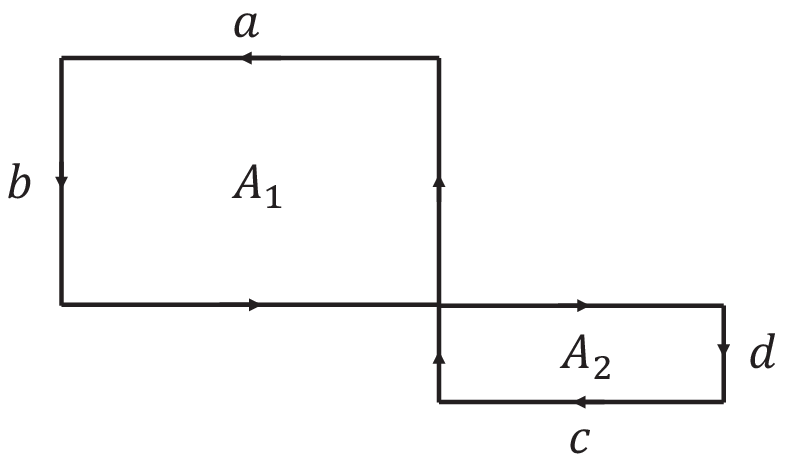}}

\caption{Schematic diagram of an $8$-loop. $A_1 = a l_0 \times b l_0$ and $A_2 = c l_0 \times d l_0$ are the scalar areas of the two rectangular loops that make up the $8$-loop.}
\label{$8$-loop}
\end{figure}

\begin{figure}
\centering
    \begin{overpic}[width=0.45\textwidth]{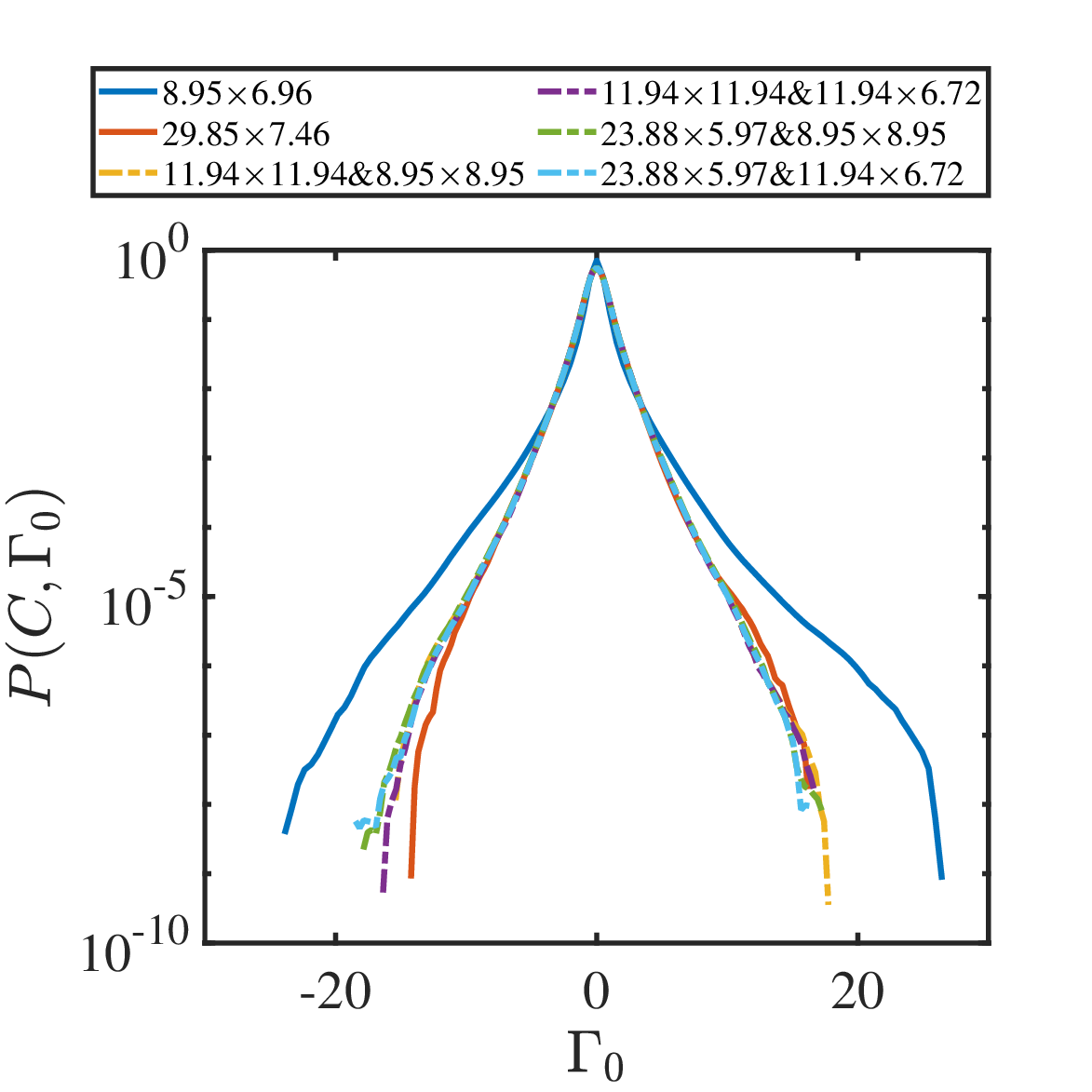}
        \put(0,75){(a)}
    \end{overpic}
    \begin{overpic}[width=0.45\textwidth]{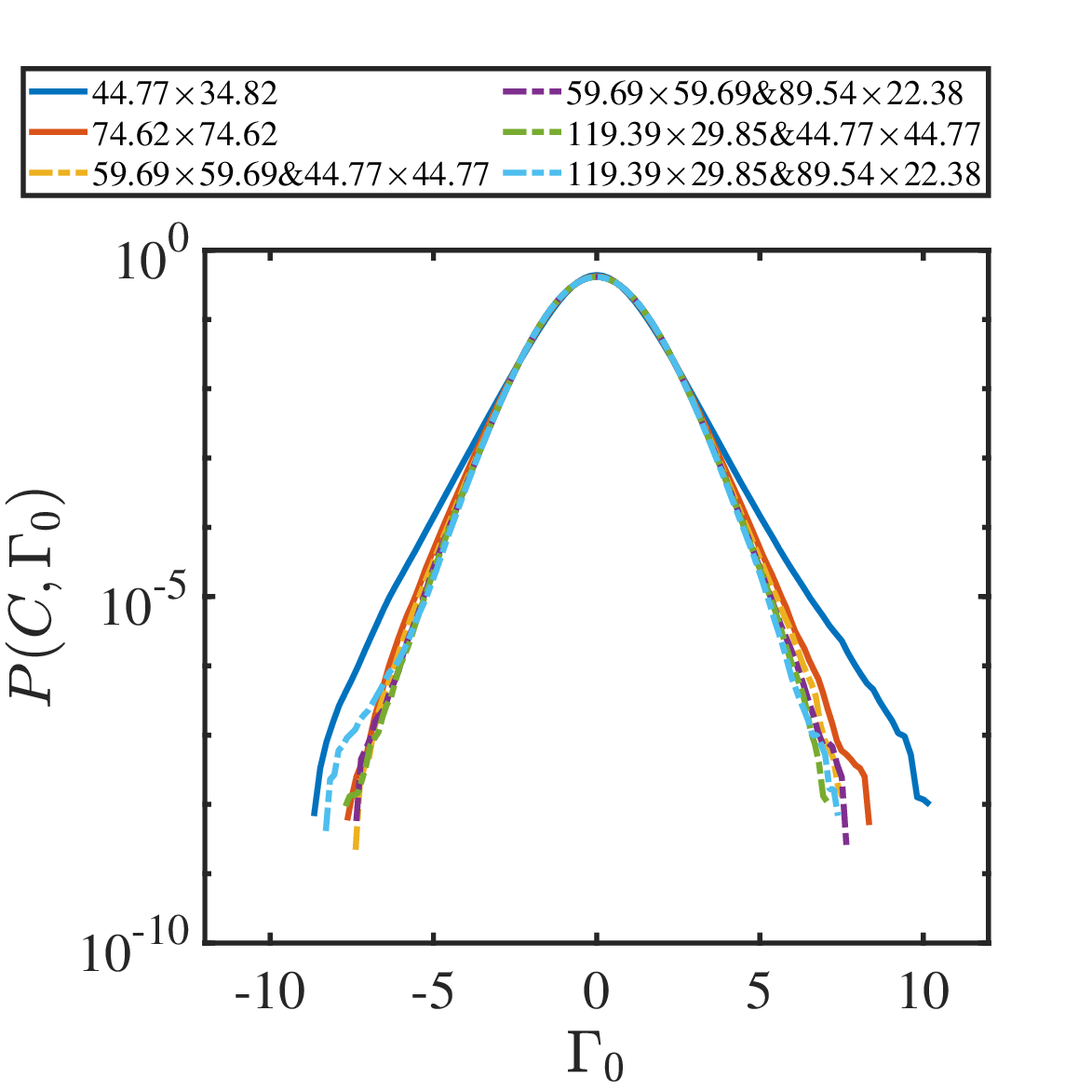}
        \put(0,75){(b)}
    \end{overpic}

\caption{Statistical results of $8$-loops. $a \times b \& c \times d$ in the legend describes the shape and size of the $8$-loop corresponding to the curve, where $a$, $b$, $c$, and $d$ are defined in Figure \ref{$8$-loop}. The blue (red) solid line corresponds to the normalized PDF of the velocity circulation on a rectangular loop with an area equal to the sum of the vector (scalar) area of 8-loops. The length of the rectangular loop is $e l_0$ and the width is $f l_0$ ($e \times f$ in the legend). $\Gamma_0 = \Gamma / \sqrt{s_\Gamma}$ is normalized circulation, where $s_\Gamma$ is the second-order moment of velocity circulation.}
\label{PDFs of 8-loops}
\end{figure}

Figure \ref{PDFs of 8-loops} shows that in the 2-D instability-driven system, the normalized PDF of the velocity circulation on an 8-loop is more determined by the scalar area sum of the 8-loops than the vector area sum. 
This result is consistent with the findings of \cite{Makoto1993} and \cite{Iyer2019}, and indicates that energy injection and dissipation may not violate the normalized area rule.

\section{Summary and discussion}
\label{Sec_diss}

For the 2-D flow in a period box, we derive the relation between the second-order moment of the filtered vorticity field and the energy spectrum, and numerically indicate that the area rule does not strictly hold in the inertial range. 
Then we derive the area rule in fully developed homogeneous instability-driven 2-D turbulence, which states that for an orientable loop, the PDF of the velocity circulation depends only on the area enclosed by the loop and is independent of the specific shape of the loop. 
By introducing an instability-driven system, we can generalize the area rule, originally derived in the inertial range \cite{Migdal2019}, to scales with energy injection and dissipation.
However, similar to 2-D turbulence with distinctive inertial ranges, numerical simulations of an instability-driven system show that the area rule does not hold strictly.  
For rectangular loops with the same enclosing area but different aspect ratios, the second-order moment of circulation varies with the aspect ratio, although their corresponding PDFs of velocity circulation have a similar shape, they cannot overlap well. 
Thus, the PDF of circulation depends mainly on the area enclosed by the loop and is influenced by the shape of the loop; there should be solutions of the loop equation other than the area rule.

Similar to \cite{Iyer2021}, we find that PDFs of loops with the same area overlap better after normalization by the second-order moment of velocity circulation. 
The results of numerical simulations suggest that $P(C,\Gamma) = \frac{1}{\sqrt{s_{\Gamma}}} P \rbr{A_C,\frac{\Gamma}{\sqrt{s_{\Gamma}}}}$ where $s_{\Gamma}$ is the second-order moment of velocity circulation on loop $C$ is a good approximate solution of the loop equation (\ref{Loop Eq, two terms}). This is also an issue for further study.

In the previous proof, we did not consider the effect of the general external force $\boldsymbol{F}$ in the equation (\ref{general momentum equation}). In fact, if and only if the external force $\boldsymbol{F}$ satisfies the condition
\begin{equation}\label{Condition of external force}
    \abr{\oint_C {\boldsymbol{F} \cdot d\boldsymbol{r}}} = 0 ,
\end{equation}
(\ref{area rule}) is the solution of the loop equation, which means the area rule can be correct. Therefore, if someone wants to test the area rule through experiments, the experimenter needs to try to make the driving force satisfy (\ref{Condition of external force}) as much as possible, or make the driving force satisfy (\ref{Condition of external force}) within a certain scale range, and then test the area rule within this scale range.

We discussed the applicability of the area rule to $8$-loops based on numerical results.
Similar to \cite{Iyer2019}, we confirmed that the area rule applies to $8$-loops and that the shape of the normalized PDF depends on the scalar sum of areas rather than the vector sum. 
The statistical properties of circulations on more general complex loops remain to be investigated. \\

\begin{acknowledgments}
The authors appreciate helpful discussions with Xi Chen, Katepalli R. Sreenivasan, Jian-Jun Tao and Hang-Yu Zhu.
We acknowledge financial support from the National Natural Science Foundation of China, grant numbers 12272006, 12472219, 42361144844 and 12588201, and from the Laoshan Laboratory under grant numbers LSKJ202202000, LSKJ202300100 and LSJKJ202400203.
\end{acknowledgments}

\appendix
\section{Taylor expansion of delta function}
\label{Sec_Appendix_A}

The delta function has the following property: 
\begin{equation} \label{delta function property}
    g^{(k)}(x_0) = (-1)^k \int_{-\infty}^{\infty} {g(x) \delta^{(k)} (x-x_0) dx} ,
\end{equation}
where $g(x)$ is a test function, $g^{(k)}(x)$ is the $k$-order derivative of $g(x)$, $\delta^{(k)}(x)$ is the $k$-order derivative of $\delta (x)$. The Taylor expansion of $g(x)$ near $x = x_0$ is
\begin{equation} \label{Taylor expansion}
    g(x_0 +\Delta x) = g(x_0) + \sum_{k=1}^{N-1} {\frac{1}{k!} g^{(k)}(x_0) (\Delta x)^k} + \frac{1}{N!} g^{(N)}(x_0 + \xi) (\Delta x)^{N} .
\end{equation}
Substitute (\ref{delta function property}) into the above equation, considering that the test function $g(x)$ is arbitrary, we have
\begin{equation} \label{delta-Taylor expansion}
    \delta (x - (x_0 +\Delta x)) = \delta (x - x_0) + \sum_{k=1}^{N-1} {\frac{(-1)^k}{k!} \delta^{(k)}(x - x_0) (\Delta x)^k} + \frac{(-1)^N}{N!} \delta^{(N)}(x - (x_0 + \xi)) (\Delta x)^{N} ,
\end{equation}
where $N>2$ is a positive integer. Because 
\begin{equation}
    (-1)^k \int_{-\infty}^{\infty} {f(x) \delta^{(k)} (x-x_1) (\Delta x)^n dx} = f^{(k)}(x_1) (\Delta x)^n \sim (\Delta x)^n
\end{equation}
for any test function $f(x)$, we can say the order of $\frac{(-1)^k}{k!} \delta^{(k)}(x - x_0) (\Delta x)^k$ is $k$. By taking the sum of all terms with orders higher than 1 in the RHS of (\ref{delta-Taylor expansion}) as $o(\Delta x)$, we can obtain
\begin{equation}
    \delta (x - (x_0 +\Delta x)) - \delta (x - x_0) = -\delta'(x - x_0) (\Delta x) + o(\Delta x) ,
\end{equation}
where $\delta'(x - x_0)$ is $\delta^{(1)}(x - x_0)$. Based on this, (\ref{2.6}) can be obtained.

\bibliography{area_rule}

\end{document}